\documentclass[twocolumn,superscriptaddress,aps,pre]{revtex4-1}
\usepackage{amssymb,amsmath}
\usepackage{bm}
\usepackage{graphicx}
\usepackage[utf8]{inputenc}
\usepackage[colorlinks,linkcolor=blue,citecolor=blue,urlcolor=blue]{hyperref}

\begin{document}

\title{Reflected fractional Brownian motion in one and higher dimensions}
\author{Thomas Vojta}
\affiliation{Department of Physics, Missouri University of Science and Technology,
Rolla, MO 65409, USA}

\author{Samuel Halladay}
\affiliation{Department of Physics, Missouri University of Science and Technology,
Rolla, MO 65409, USA}

\author{Sarah Skinner}
\affiliation{Department of Physics, Missouri University of Science and Technology,
Rolla, MO 65409, USA}

\author{Skirmantas Janu\v{s}onis}
\affiliation{Department of Psychological and Brain Sciences, University of California,
Santa Barbara, Santa Barbara, CA 93106, USA}

\author{Tobias Guggenberger}
\affiliation{Institute of Physics and Astronomy, University of Potsdam, D-14476
Potsdam-Golm, Germany}

\author{Ralf Metzler}
\affiliation{Institute of Physics and Astronomy, University of Potsdam, D-14476
Potsdam-Golm, Germany}

\begin{abstract}
Fractional Brownian motion (FBM), a non-Markovian self-similar Gaussian stochastic
process with long-ranged correlations, represents a widely applied, paradigmatic
mathematical model of anomalous diffusion. We report the results of large-scale
computer simulations of FBM in one, two, and three dimensions in the presence of
reflecting boundaries that confine the motion to finite regions in space.
Generalizing earlier results for finite and semi-infinite one-dimensional intervals,
we observe that the interplay between the long-time correlations of FBM and the
reflecting boundaries leads to striking deviations of the stationary probability
density from the uniform density found for normal diffusion. Particles accumulate
at the boundaries for superdiffusive FBM while their density is depleted at the
boundaries for subdiffusion. Specifically, the probability density $P$ develops a
power-law singularity, $P\sim r^\kappa$, as a function of the distance $r$ from the
wall. We determine the exponent $\kappa$ as a function of the dimensionality, the
confining geometry, and the anomalous diffusion exponent $\alpha$ of the FBM. We
also discuss implications of our results, including an application to modeling
serotonergic fiber density patterns in vertebrate brains.
\end{abstract}

\date{\today}

\maketitle

\section{Introduction}
\label{sec:Intro}

Following pioneering works of Einstein \cite{Einstein_book56}, Smoluchowski
\cite{Smoluchowski17}, and Langevin \cite{Langevin08}, normal diffusion can be understood as
random motion that is local in time and space. This means that normal diffusion is
a stochastic process that fulfills two conditions, (i) it features a finite
correlation time after which individual steps become statistically independent, and
(ii) the displacements over a correlation time feature a finite second moment. If
these conditions are fulfilled, the central limit theorem applies, resulting in
the well-known linear relation $\langle \mathbf r^2 \rangle \sim t$ between the
mean-square displacement of the moving particle and the elapsed time $t$
\cite{Hughes95}.

If at least one of the preconditions for the central limit theorem is violated, deviations from
the linear relation $\langle \mathbf r^2 \rangle \sim t$ may appear, giving rise to anomalous diffusion
(for reviews see, e.g., Refs.\
\cite{MetzlerKlafter00,HoeflingFranosch13,BressloffNewby13,MJCB14,MerozSokolov15,MetzlerJeonCherstvy16}
and references therein).
For example, sufficiently broad distributions of waiting times between individual steps can lead
to subdiffusive motion (for which $\langle \mathbf r^2 \rangle$ increases slower than $t$)
while broad distributions of step sizes may produce superdiffusion (where $\langle \mathbf r^2 \rangle$
increases faster than $t$). Anomalous diffusion is often characterized by the power-law dependence
\begin{equation}
\langle \mathbf r^2 \rangle \sim t^\alpha
\label{eq:anomalous}
\end{equation}
where $\alpha$ is the anomalous diffusion exponent which takes values $1 < \alpha < 2$ for superdiffusion
and $0 < \alpha < 1$ for subdiffusion.

Another important mechanism leading to anomalous diffusion consists of long-range correlations in time
between the displacements of the particle. The prototypical mathematical model of a stochastic process with
long-time correlated steps is fractional Brownian motion (FBM) which was introduced by Kolmogorov
\cite{Kolmogorov40} and further studied by Mandelbrot and van Ness \cite{MandelbrotVanNess68}.
FBM is a self-similar Gaussian stochastic process with stationary long-time correlated increments.
It gives rise to power-law anomalous diffusion (\ref{eq:anomalous}). In the superdiffusive regime,
$1 < \alpha < 2$, the motion is persistent (positive correlations between the steps) whereas it
is anti-persistent (negative correlations) in the subdiffusive regime, $0 < \alpha < 1$. In the marginal
case $\alpha=1$, FBM is identical to normal Brownian motion with uncorrelated steps.

FBM has been applied to model the dynamics in a wide variety of systems including diffusion inside
biological cells
\cite{SzymanskiWeiss09,MWBK09,WeberSpakowitzTheriot10,Jeonetal11,JMJM12,Tabeietal13}, the dynamics of
polymers \cite{ChakravartiSebastian97,Panja10}, electronic network traffic \cite{MRRS02}, as well as
fluctuations of financial markets \cite{ComteRenault98,RostekSchoebel13}.
FBM has been analyzed quite extensively in the mathematical literature (see, e.g., Refs.\
\cite{Kahane85,Yaglom87,Beran94,BHOZ08}) but only few results are available for FBM in confined
geometries, i.e., in the presence of nontrivial boundary conditions. These include the solution of the
first-passage problem of FBM confined to a semi-infinite interval \cite{HansenEngoyMaloy94,DingYang95,KKMCBS97,Molchan99}),
a conjecture for a two-dimensional wedge domain \cite{JeonChechkinMetzler11},
and corresponding results for parabolic domains \cite{AurzadaLifshits19}. In addition, the
probability density of FBM on a semi-infinite interval with an absorbing boundary was
investigated in Refs.\ \cite{ChatelainKantorKardar08,ZoiaRossoMajumdar09,WieseMahumdarRosso11}.
The difficulties in analyzing FBM in confined geometries are related to the fact that a generalized diffusion
equation for FBM applicable to solve boundary value problems is yet to be found,
and the method of images \cite{MetzlerKlafter00,Redner_book01}, typically invoked for
boundary value problems, fails.

Recently, FBM with reflecting walls has attracted considerable attention as computer simulations
have demonstrated that
the interplay between the long-time correlations and the confinement modifies the probability density
function $P(x,t)$ of the diffusing particles. For FBM on a semi-infinite
interval with a reflecting wall at the origin, the probability density becomes highly non-Gaussian and develops
a power-law singularity, $P \sim x^\kappa$, at the wall \cite{WadaVojta18,WadaWarhoverVojta19}.
For persistent noise (superdiffusive FBM), particles accumulate at the wall, $\kappa<0$, whereas particles are depleted near
the wall, $\kappa > 0$ for anti-persistent noise (subdiffusive FBM). Analogous simulations of FBM
on a finite interval, with reflecting walls at both ends, have shown that the stationary probability density
deviates from the uniform distribution found for normal diffusion
\cite{Guggenbergeretal19}. Particles accumulate at the walls and are depleted in the middle of the interval
for persistent noise whereas the opposite is true for anti-persistent noise.

The above results for the probability density of reflected FBM are all restricted to one dimension whereas
many of the applications in physics, biology and beyond are in two or three dimensions. It is therefore
interesting and important to ask whether reflected FBM in higher dimensions also features unusual
accumulation and depletion effects of particles near reflecting boundaries and to determine the
functional form of the probability density in these cases.

In the present paper, we therefore analyze by means of large-scale computer simulations the properties
of reflected FBM in various confined geometries. After providing some additional results in one dimension,
the main focus will be on reflected FBM in two and three space dimensions. In all cases, we find that
particles accumulate at the reflecting walls for persistent noise and are depleted close to the walls
for anti-persistent noise, just as in one dimension.  The probability density behaves as a power of
the distance from the wall, $P \sim r^\kappa$. We determine the exponent $\kappa$ as a function of the
dimensionality, the confining geometry, and the anomalous diffusion exponent $\alpha$ of the FBM.

Our paper is organized as follows. We define reflected FBM in one and higher dimensions in Sec.\
\ref{sec:FBM} where we also discuss the details of our numerical approach. Sections \ref{sec:1d},
\ref{sec:2d}, and \ref{sec:3d} are devoted to results for one, two, and three space dimensions,
respectively. In Sec.\ \ref{sec:fibers}, we discuss an interesting application of reflected FBM
to model serotonergic fibers in vertebrate brains
\cite{JanusonisDetering19,JanusonisDeteringMetzlerVojta20}.
We conclude in Sec.\ \ref{sec:conclusions}.

\section{Reflected fractional Brownian motion}
\label{sec:FBM}
\subsection{Definition of FBM}
\label{subsec:Def_FBM}

We start by defining FBM in one space dimension. FBM is a continuous-time centered Gaussian stochastic
process. The covariance function of the position $X$ at times $s$ and $t$ is given by
\begin{equation}
\langle X(s) X(t) \rangle = K (s^\alpha - |s-t|^\alpha + t^\alpha)
\label{eq:FBM_cov}
\end{equation}
defined for anomalous diffusion exponents $\alpha$ in the range $0 < \alpha < 2$.
Setting $s=t$, this yields anomalous diffusion with a mean-square displacement of
$\langle X^2 \rangle = 2 K t^\alpha$, i.e., superdiffusion for $\alpha>1$
and subdiffusion for $\alpha<1$.
Correspondingly, the probability density of unconfined (free space) FBM takes the Gaussian form
\begin{equation}
	P(x,t) = \frac{1}{\sqrt{4\pi K t^\alpha}} \exp{ \left( -\frac{x^2}{4 K t^\alpha} \right) }~.
\label{eq:FBM_P(x)_free}
\end{equation}
We now discretize time by defining $x_n = X(t_n)$ with $t_n= \epsilon n$ where
$\epsilon$ is the time step and $n$ is an integer. This leads to a discrete version
of FBM \cite{Qian03} that lends itself to computer simulations. It can be
understood as a random walk with identically Gaussian distributed but long-time
correlated steps. Specifically, the position $x_n$ of the particle evolves according
to the recursion relation
\begin{equation}
x_{n+1} = x_n + \xi_n~.
\label{eq:FBM_recursion}
\end{equation}
The increments $\xi_n$ are a discrete fractional Gaussian noise, a stationary
Gaussian process of zero mean, variance $\sigma^2 = 2 K \epsilon^\alpha$, and covariance function
\begin{equation}
C_n=\langle \xi_m \xi_{m+n} \rangle = \frac 1 2 \sigma^2 (|n+1|^\alpha - 2|n|^\alpha + |n-1|^\alpha)~.
\label{eq:FGN_cov}
\end{equation}
The correlations are positive (persistent) for $\alpha>1$ and negative (anti-persistent)
for $\alpha < 1$. In the marginal case, $\alpha=1$, the covariance vanishes for all
$n\ne 0$, i.e., we recover normal Brownian motion. For $n\to \infty$, the covariance takes the
power-law form $\langle \xi_m \xi_{m+n} \rangle  \sim\alpha (\alpha-1) |n|^{\alpha-2}$.

To reach the continuum limit, the time step $\epsilon$ needs to be small compared to the considered
times $t$. Equivalently, the size $\sigma$ of an individual increment must be small compared to the
considered distances or system sizes.
This can be achieved either by taking $\epsilon$ to zero at fixed $t$ or, equivalently,
by taking $t$ to infinity at fixed $\epsilon$. In this paper, we chose the latter route by fixing
$\epsilon=\mathrm{const}$ and considering times $t \to \infty$.

We now generalize FBM from one to higher dimension. FBM in $d$ dimensions can be defined as
the superposition of $d$ independent FBM processes, one for each Cartesian coordinate
\cite{QianRaymondBassingthaingthe98,JeonMetzler10}. This means the $d$-dimensional position
vector $\mathbf{r}_n$ follows the recursion relation
\begin{equation}
\mathbf{r}_{n+1} = \mathbf{r}_n + \bm{\xi}_n
\label{eq:FBM_recursion_d}
\end{equation}
where the components $\xi_n^{(i)}$ of the $d$-dimensional fractional
Gaussian noise feature the covariance function
\begin{equation}
\langle \xi_m^{(i)} \xi_{m+n}^{(j)} \rangle = \frac 1 2 \sigma^2 (|n+1|^\alpha - 2|n|^\alpha + |n-1|^\alpha) \delta_{ij}~.
\label{eq:FGN_cov_d}
\end{equation}
It is easy to show that this definition is invariant under rotations of the coordinate system.
We also note that the generalization of FBM to higher dimensions as superposition of
independent components is not unique. More complicated correlation structures between the
components have been considered in the mathematical literature (see, e.g., Ref.\ \cite{ACLP13}).

\subsection{Reflecting boundaries}
\label{subsec:wall_def}

Let us now discuss how to define the boundary conditions that confine the FBM to a given geometry.
Reflecting walls can be implemented by suitably modifying the recursion relations
(\ref{eq:FBM_recursion}) and (\ref{eq:FBM_recursion_d}). As the fractional Gaussian noise is understood
as externally given \cite{Klimontovich_book95}, it is not affected by the walls.
In one dimension, an ``elastic'' wall at position $w$ that restricts the motion to $x \ge w$
(i.e., a wall to the left of the allowed interval) can be defined by means of
\begin{equation}
x_{n+1} = w + | x_n + \xi_n - w |~.
\label{eq:FBM_recursion_elastic}
\end{equation}
This definition was employed in recent studies of reflected FBM
\cite{JeonMetzler10,WadaVojta18,WadaWarhoverVojta19,Guggenbergeretal19}, but it is by no means unique.
The recursion relation
\begin{equation}
x_{n+1} = \left \{ \begin{array}{ll} x_n+ \xi_n  & \quad \textrm{if} \quad x_n+ \xi_n \ge w\\
                     x_n & \quad \textrm{otherwise}   \end{array}  \right.
\label{eq:FBM_recursion_inelastic}
\end{equation}
defines an ``inelastic'' wall at which the particle does not move at all if the step would take it
into the forbidden region $x < w$. Alternatively, the recursion
\begin{equation}
x_{n+1} = \max(x_n+ \xi_n,w)
\label{eq:FBM_recursion_stuck}
\end{equation}
places the particle right at the wall if the step would take it into the forbidden region $x < w$.
Definition (\ref{eq:FBM_recursion_stuck}) can be understood as a discretized version
of the definition of reflected FBM in the mathematical literature where it is employed,
e.g., in queueing theory \cite{Harrison85,Whitt02}.

In addition to these hard walls one can also introduce soft walls by adding repulsive forces to
the recursion relation,
\begin{equation}
x_{n+1} =  x_n + \xi_n + F(x_n)~.
\label{eq:FBM_recursion_soft}
\end{equation}
We consider exponential forces,
\begin{equation}
F(x) = F_0 \exp[- \lambda (x-w)]~,
\label{eq:wall_force}
\end{equation}
characterized by amplitude $F_0$ and decay constant $\lambda$ . Note that a factor $\epsilon$ stemming from the time step
has been absorbed in the amplitude $F_0$.
Boundaries restricting the motion to positions $x \le w$ (i.e., walls at the right end of an
allowed interval) can be defined in analogy to (\ref{eq:FBM_recursion_elastic}) to (\ref{eq:FBM_recursion_soft}).

In higher dimensions, we use appropriate generalizations of the wall implementations
(\ref{eq:FBM_recursion_elastic}), (\ref{eq:FBM_recursion_inelastic}), and
(\ref{eq:FBM_recursion_soft}). This is unambiguous for the ``inelastic'' wall
which prevents the particle from moving if it would enter the forbidden region,
\begin{equation}
\mathbf{r}_{n+1} = \left \{ \begin{array}{ll}
        \mathbf{r}_n + \bm{\xi}_n  & \quad \textrm{if} ~ \mathbf{r}_n + \bm{\xi}_n ~\textrm{is in allowed region}\\
                     \mathbf{r}_n & \quad \textrm{otherwise}   \end{array}  \right.~.
\label{eq:FBM_recursion_inelastic_d}
\end{equation}
For other wall implementations, some care is required to properly deal with the
directions of the motion and of the wall forces, in particular in complex geometries.
For example, a simple reflection analogous to (\ref{eq:FBM_recursion_elastic}) becomes
ambiguous if the allowed region features sharp corners, and, unless the geometry is highly
symmetric, the directions of the wall forces depend on details of the modeling potential.

In the following, the majority of our simulations utilize the ``inelastic'' walls
(\ref{eq:FBM_recursion_inelastic}) and (\ref{eq:FBM_recursion_inelastic_d}).
However, for reflected FBM to be a well-defined self-contained concept, it is important to
establish that its properties do not depend on the precise choice of boundary conditions
(so that it can be applied to situations in which details of the interactions between
the particles and the wall are not known). In Sec.\ \ref{subsec:wall_test}, we therefore
carefully compare trajectories and probability densities resulting from different wall
implementations. The data show that the wall implementations affect the immediate vicinity
of wall only and become unimportant in the continuum limit, i.e., on length scales large
compared to $\sigma$ and $\lambda^{-1}$.

\subsection{Simulation details}

In the following sections, we report results of computer simulations of our discrete-time
FBM in one, two, and three dimensions for anomalous diffusion exponents $\alpha$ in the range
between 0.3 (deep in the subdiffusive regime) and 1.95 (deep in the superdiffusive regime and
almost at the ballistic limit $\alpha=2$). Each simulation uses a large number of particles,
up to $10^7$. We fix the time step at $\epsilon=1$ and set $K=1/2$ (unless noted otherwise).
This implies a variance $\sigma^2=1$ of the individual steps. Each particle performs
up to $2^{29} \approx 5.4 \times 10^8$ time steps.

As discussed in Sec.\ \ref{subsec:Def_FBM}, this large number of steps allows us to reach the
continuum (scaling) limit for which the time discretization becomes unimportant, and the behavior
approaches that of continuous time FBM. Expressed in terms of the linear system size $L$,
the continuum limit takes the form $L/\sigma \gg 1$. In our simulations, the linear system sizes
range from $L=100$ for the most
subdiffusive $\alpha=0.3$ to $L=10^6$ for some calculations using $\alpha$ values close to 2.

The correlated Gaussian random numbers $\xi_n$  that represent the fractional noise are
precalculated before each actual simulation by means of the Fourier-filtering technique \cite{MHSS96}.
For each Cartesian component of the noise, this method starts from a sequence of independent
Gaussian random numbers $\chi_i$ of zero mean and unit variance (which we generate using the Box-Muller transformation with the LFSR113 random number generator proposed by L'Ecuyer \cite{Lecuyer99} as well as
the 2005 version of Marsaglia's KISS \cite{Marsaglia05}).
The Fourier transform $\tilde \chi_\omega$ of these numbers
is then converted via ${\tilde{\xi}_\omega} = [\tilde C(\omega)]^{1/2} \tilde{\chi}_\omega$,
where $\tilde C(\omega)$ is the Fourier transform of the covariance function (\ref{eq:FGN_cov}).
The inverse Fourier transformation of the ${\tilde{\xi}_\omega}$ gives the desired noise values.

\section{One space dimension}
\label{sec:1d}
\subsection{Summary of earlier results}

Wada et al.\ \cite{WadaVojta18,WadaWarhoverVojta19} recently employed computer simulations to
study one-dimensional FBM restricted to the semi-infinite interval $(0,\infty)$ by a reflecting wall
at the origin. They observed that the mean-square displacement $\langle x^2 \rangle$ of a particle
that starts at the origin follows the expected power law $t^\alpha$ just as for unconfined FBM.
However, the probability density was found to be highly non-Gaussian with particles accumulating
at the wall in the superdiffusive regime $\alpha>1$. For subdiffusive FBM, $\alpha < 1$, particles
are depleted near the wall.

More specifically, the probability density function $P(x,t)$ of the particle position $x$ at time $t$
fulfills the scaling form
\begin{equation}
P(x,t) = \frac 1 {\sigma t^{\alpha/2}}\, Z_\alpha \left[ x/(\sigma t^{\alpha/2}) \right ]
\label{eq:P_scaling_t}
\end{equation}
in the continuum limit $x \gg \sigma$. The dimensionless scaling function $Z_\alpha(z)$ is non-Gaussian
near the wall; it develops a singularity for $z \to 0$. Based on the extensive simulation results,
Wada et al.\ conjectured a power-law singularity $Z_\alpha(z) \sim z^\kappa$ for $z \ll 1$ with
the exponent given by $\kappa =2/\alpha -2$.

Analogous results were also obtained for biased FBM on the interval $(0,\infty)$
\cite{WadaWarhoverVojta19}. If the bias is towards
the wall, a stationary distribution develops in the long time limit. Its probability density
also features a power-law singularity at the wall, $P(x) \sim x^\kappa$, controlled by the same
exponent $\kappa=2/\alpha-2$.

Guggenberger et al.\ \cite{Guggenbergeretal19} performed simulations of one-dimensional
FBM confined to a finite interval by reflecting walls at both ends. They established that,
for all $\alpha \ne 1$, the stationary probability density $P(x)$ deviates from the uniform
distribution observed for normal Brownian motion. For $\alpha > 1$, the probability density
is increased at the walls and reduced in the middle of the interval. For $\alpha < 1$,
the opposite behavior is observed. However, the functional form of the probability density
on a finite interval has not yet been studied systematically.

In the rest of this section, we therefore analyze one-dimensional FBM on long intervals of
lengths $L \gg \sigma$, with reflecting walls at both ends. We use up to $2^{29}$ time steps.
This allows us to determine the probability density and, in particular,
analyze its functional form close to the walls. In addition, we carefully study the effects
of different wall implementation on the probability density.

\subsection{Reflected FBM on a finite interval}

\begin{figure}
\includegraphics[width=\columnwidth]{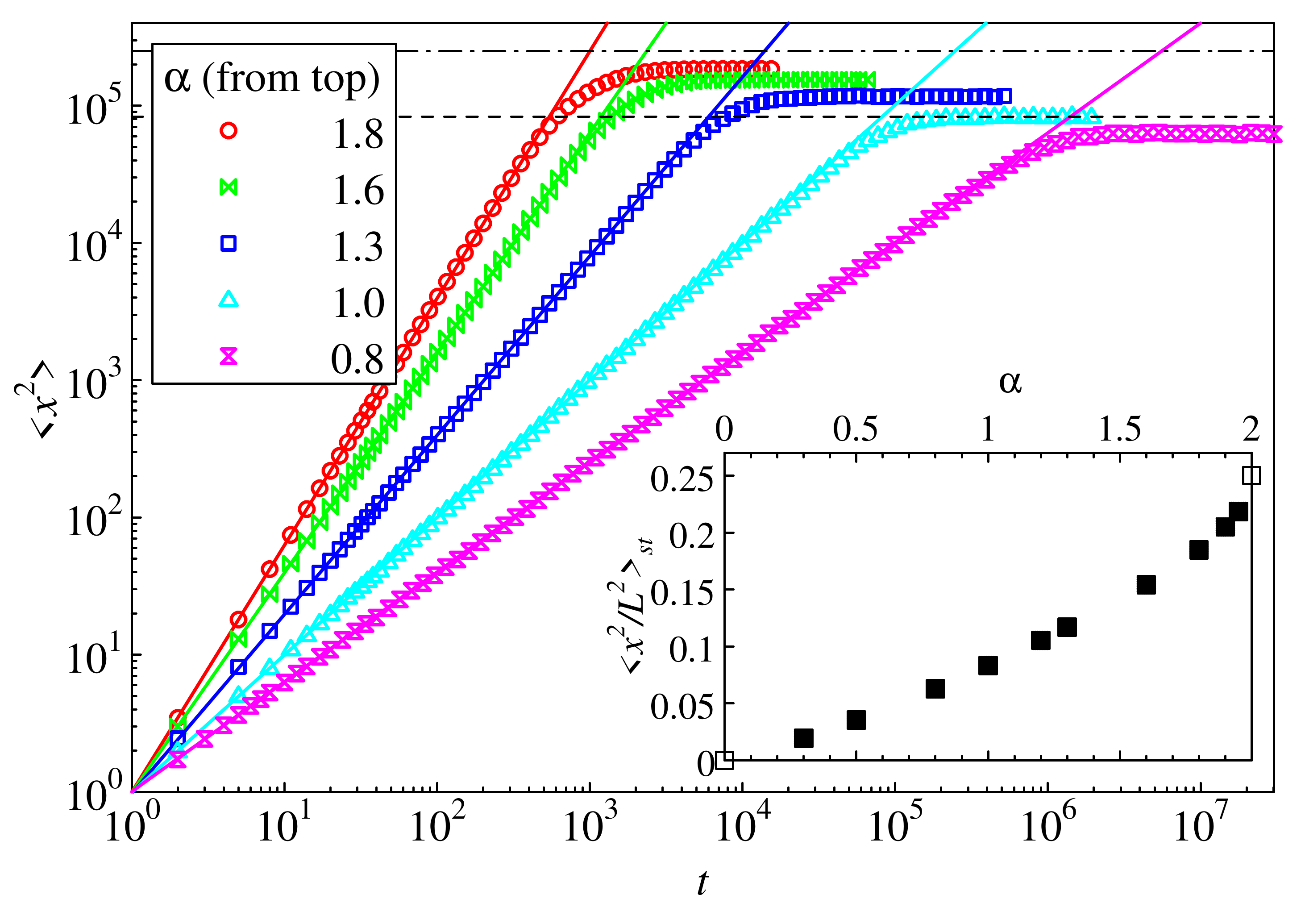}
\caption{Mean square displacement $\langle x^2 \rangle$ on the
interval $(-L/2,L/2)$ with $L=1000$ for several $\alpha$. The data are averages
over 10,000 particles starting at $x=0$. The reflecting walls are implemented
using eq.\ (\ref{eq:FBM_recursion_inelastic}). The dashed line marks the value
$L^2/12\approx 83,333$ expected for a uniform distribution of particles over the
interval, and the dash-dotted line marks the value $L^2/4 = 250,000$ expected if all particles
 collect at the walls. The solid lines are fits of the initial time evolution to
$\langle x^2 \rangle \sim t^\alpha$. Inset: Normalized stationary mean-square displacement
$A(\alpha) =\langle x^2 \rangle_{\mathrm{st}}/L^2$
vs.\ the anomalous diffusion exponent $\alpha$ using interval lengths
between $L=100$ for $\alpha=0.3$ and $L\ge 10,000$ for the largest $\alpha$. The open
squares mark the values expected in the limits $\alpha \to 0$ and $\alpha \to 2$.
The statistical errors of $\langle x^2 \rangle_{\mathrm{st}}/L^2$ are much smaller than the symbol size.  }
\label{fig:1d_MSD}
\end{figure}

We first study the time evolution of the mean-square displacement $\langle x^2 \rangle$ of FBM
on the interval $(-L/2, L/2)$. The particles start at the origin, i.e., in the center of the
interval. Figure \ref{fig:1d_MSD} presents the mean-square displacement for an interval of
length $L=1000$ for several different anomalous diffusion exponents $\alpha$.
The figure demonstrates that $\langle x^2 \rangle$ initially grows following the same $t^\alpha$
power law as unconfined FBM. At long times it saturates at a stationary value $\langle x^2 \rangle_{\mathrm{st}}$
that changes with $\alpha$,
suggesting a nonuniform and $\alpha$-dependent distribution of particles in the stationary state.
In the continuum limit $L \gg \sigma$, the stationary mean-square displacement is proportional to $L^2$.
 (This also follows from the scaling law (\ref{eq:P_scaling_L}) discussed below.)
The inset of Fig.\ \ref{fig:1d_MSD} indicates that $A(\alpha) = \langle x^2 \rangle_{\mathrm{st}} / L^2$
evolves smoothly with $\alpha$ from the value 1/4 expected in the ballistic limit $\alpha \to 2$
(where all particles get stuck directly at the walls) to the value 0 for $\alpha \to 0$
(where the particles do no leave the center). The crossover time $t_x$ between anomalous diffusion and
saturation follows from $\sigma^2 t_x^\alpha = A(\alpha) L^2$.

We emphasize that the functional behavior of the stationary mean-square displacement
shown here is strikingly different from the one obtained for FBM in a harmonic
confining potential, represented by a force $F(x) = - b x$ in Eq.\ (\ref{eq:FBM_recursion_soft}).
(Note that FBM does not fulfill a fluctuation-dissipation relation and is thus not thermalized.)
There, the mean-square displacement takes the value $\langle x^2\rangle_{
\mathrm{st}}^{\mathrm{harm}}=\frac{1}{2}\sigma^2 b^{-\alpha} \Gamma(\alpha+1)$ \cite{Sliusarenko10}.
In particular, the value of $\langle x^2\rangle_{\mathrm{st}}^{\mathrm{harm}}/(
\sigma^2 b^{-\alpha}/2)$ is unity for $\alpha \to 0$ and at $\alpha=1$, attains its minimum of
about $0.89$ at $\alpha\approx 0.46$, and reaches its maximum of 2 in the ballistic limit
$\alpha \to 2$
\footnote{These relations hold in the continuum limit $\protect\langle x^2\protect\rangle_{
\mathrm{st}}^{\mathrm{harm}} \gg \sigma^2$.}.

\begin{figure}
\includegraphics[width=\columnwidth]{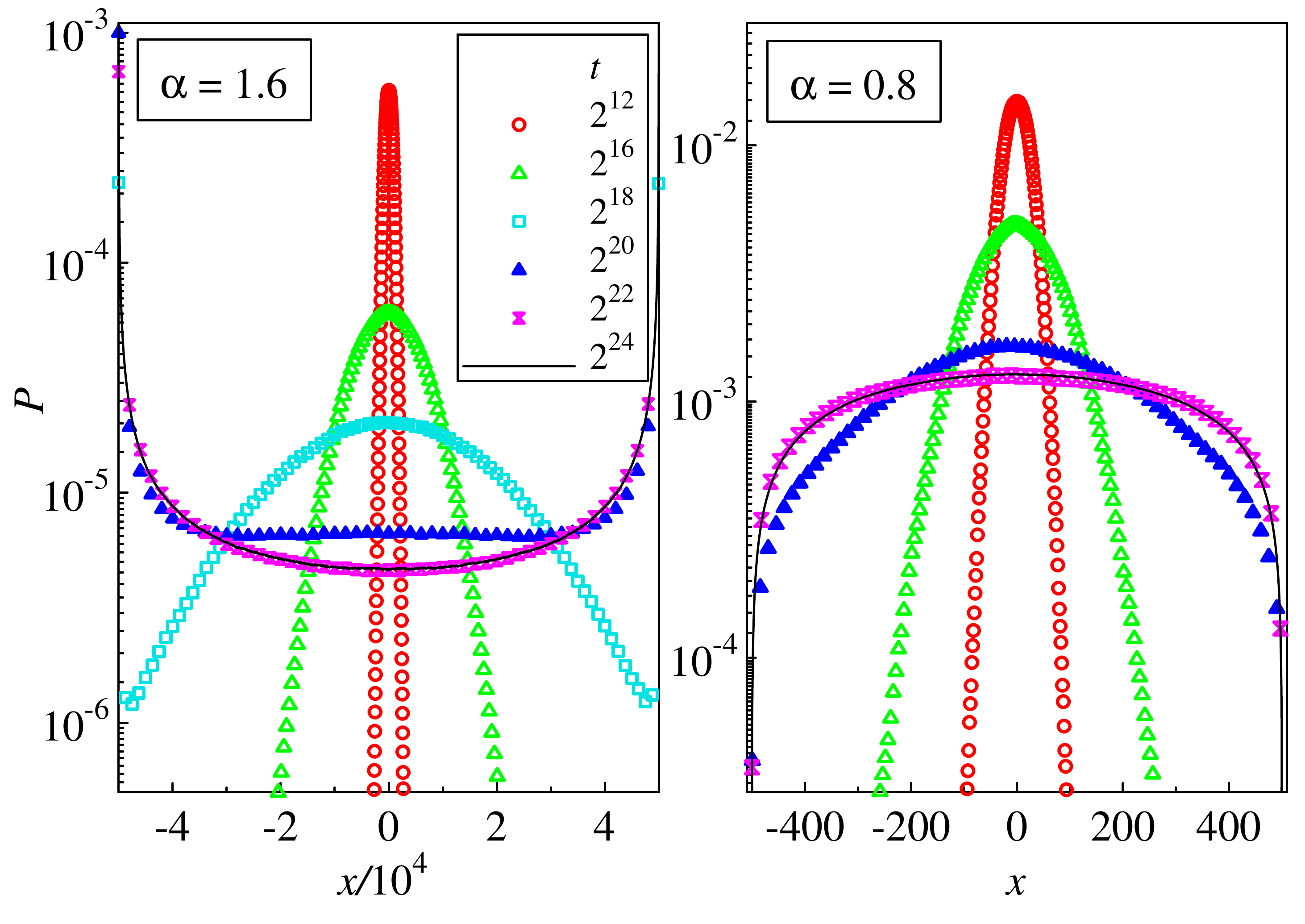}
\caption{Log-linear plots of the probability density $P$ vs.\ position $x$ for $\alpha=1.6$ and 0.8
at several times $t$. The particles start at the center of intervals of length $L=10^5$ and $10^3$,
respectively.
Each distribution is based on at least $10^5$ particles. To improve
the statistics, $P$ is averaged over a small time interval around each of the given times.
The statistical errors of $P$ are smaller than the symbol size.}
\label{fig:1d_PDF_gam04}
\end{figure}

We now turn to the time evolution of the probability density function $P(x,t)$.
Figure \ref{fig:1d_PDF_gam04} shows the probability density for $\alpha=1.6$ and 0.8 at several
different times.
At early times, it is not affected by the walls and takes a Gaussian form, just as for free FBM.
Once the distribution interacts with the reflecting walls, particles start to accumulate close to the walls
in the superdiffusive case $\alpha=1.6$ while $P$ remains suppressed at the walls for the subdiffusive
case $\alpha=0.8$.
The probability density reaches a non-uniform stationary state for times larger than approximately
$2^{22}$.

\begin{figure}
\includegraphics[width=\columnwidth]{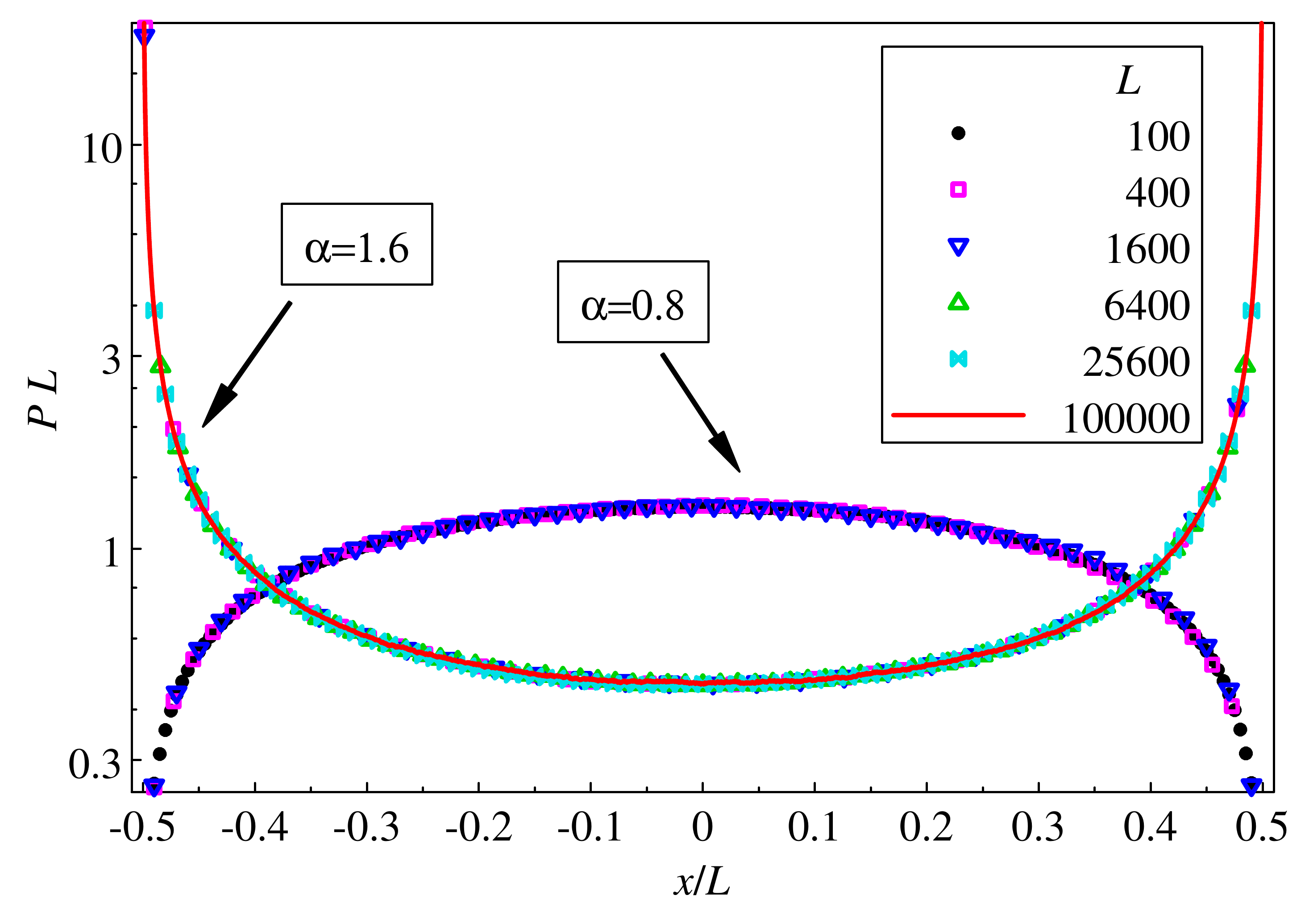}
\caption{Scaling plot of the stationary probability density showing $P L$ vs.\ $x/L$
for $\alpha=1.6$ and $\alpha=0.8$ for several interval length $L$. Each distribution is based
on $10^4$ to $10^5$ particles; $P$ is averaged over a number of time steps after a stationary state
has been reached.}
\label{fig:1d_PDF_gam04_scaled}
\end{figure}

It is interesting to compare the form of the stationary probability density $P$ for different
interval lengths $L$. Figure \ref{fig:1d_PDF_gam04_scaled} presents the corresponding simulation data
for several $L$ between 100 and $10^5$ using scaled variables $P L$ vs.\ $x/L$.
The curves for different $L$ collapse nearly perfectly onto a common master curve, demonstrating
that the stationary distribution fulfills the scaling form
\begin{equation}
P(x,L) = \frac 1 L Y_\alpha(x/L)
\label{eq:P_scaling_L}
\end{equation}
with high accuracy. Small deviations (almost invisible in the figure) can be attributed to finite-size
effects close to the wall that vanish in the continuum limit $L \gg \sigma$. We have carried out similar
simulations for other values of the anomalous diffusion exponent $\alpha$. All data fulfill the scaling
relation (\ref{eq:P_scaling_L}) but the functional form of the dimensionless scaling function $Y_\alpha(z)$
depends on $\alpha$. Note that the scaling form (\ref{eq:P_scaling_L}) also implies that
the probability density of FBM on an interval of fixed length becomes independent of the step size $\sigma$
in the limit $\sigma \to 0$, guaranteeing that our reflected FBM has a proper continuum limit.

\begin{figure}
\includegraphics[width=\columnwidth]{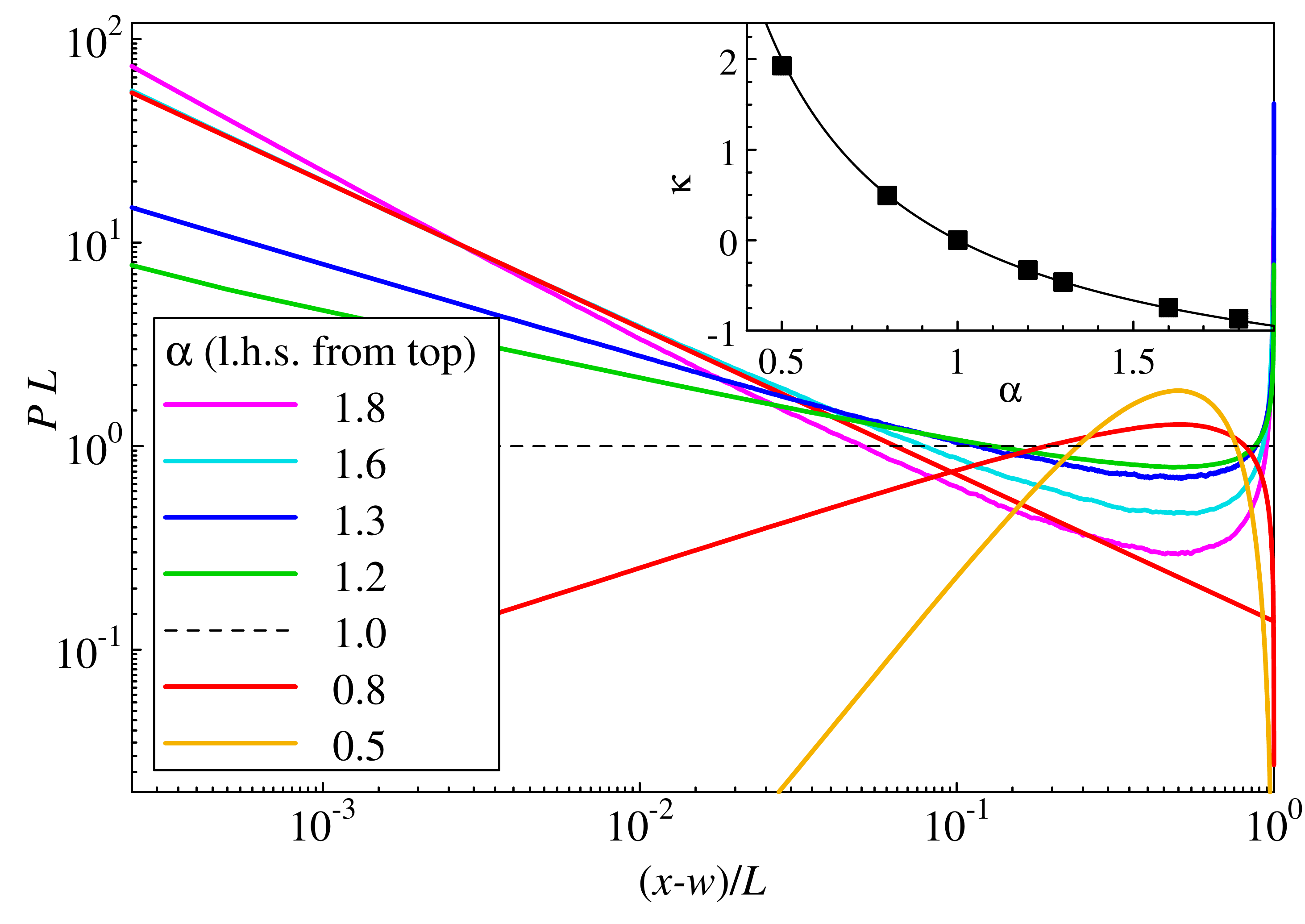}
\caption{Scaled stationary probability density $P L$ vs.\ scaled distance $(x-w)/L$ from the
wall for several $\alpha$. The system sizes range from $L=200$ for $\alpha=0.5$ to $L=10^6$
for the largest $\alpha$. Each distribution is based on $10^4$ to $10^5$ particles;
$P$ is averaged over a large number (up to $2^{25}$) of time steps after a stationary state
has been reached.   Inset: Exponent $\kappa$ extracted from power-law fits of $P(x)$ close to
the wall. The solid line is the  conjecture $\kappa =2/\alpha -2$. }
\label{fig:1d_PDF_log_scaled_all}
\end{figure}

Let us now focus on the behavior of the stationary probability density close to the reflecting walls.
Based on the results for reflected FBM on a semi-infinite interval \cite{WadaVojta18,WadaWarhoverVojta19},
we expect the probability density to feature a power-law singularity at the wall.
In Fig.\ \ref{fig:1d_PDF_log_scaled_all}, we therefore present a double-logarithmic plot of the probability density
as a function of the distance from the reflecting wall for several $\alpha$ between 0.5 and 1.8.
The figure shows that all curves become straight lines close to the wall, indicating that the
stationary probability density indeed follows the power-law $P \sim (x-w)^\kappa$. We determine the values
of the exponent $\kappa$ by power-law fits of the probability density close to the wall but outside
of the region influenced by finite-size effects, i.e., for $\sigma \ll x-w \ll L$.
The inset of Fig.\ \ref{fig:1d_PDF_log_scaled_all} shows $\kappa$ as a function of $\alpha$.
The exponent follows the conjecture $\kappa =2/\alpha -2$ with high accuracy, i.e., it takes
the same values as the exponent in the case of a semi-infinite interval. This implies that the scaling
function $Y_\alpha$ in Eq. (\ref{eq:P_scaling_L}) behaves as $Y_\alpha(z) \sim (z+1/2)^\kappa =
(z+1/2)^{2/\alpha -2}$
for $z+1/2 \ll 1$ (close to the left interval boundary) and analogously for the right boundary.

\subsection{Influence of wall implementation}
\label{subsec:wall_test}

\begin{figure}
\includegraphics[width=\columnwidth]{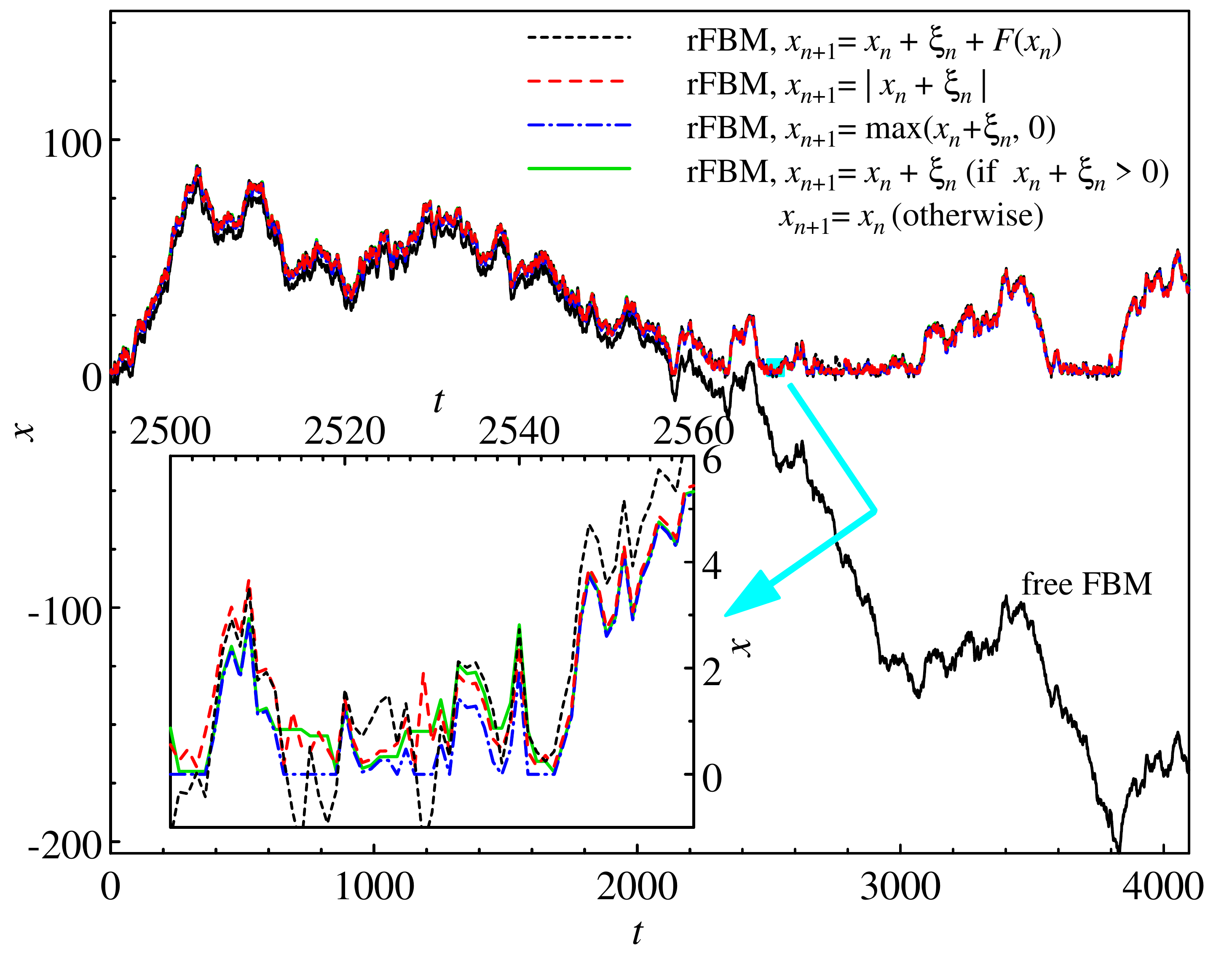}
\caption{Example trajectories of FBM ($\alpha=1.2$) with a reflecting wall at the origin (rFBM), implemented via the
boundary conditions (\ref{eq:FBM_recursion_elastic}), (\ref{eq:FBM_recursion_inelastic}),
(\ref{eq:FBM_recursion_stuck}), and (\ref{eq:FBM_recursion_soft}). For the soft wall (\ref{eq:FBM_recursion_soft}),
the force parameters are $F_0=0.5$ and $\lambda=1$. The free FBM trajectory is shown for comparison.
Inset: Zoomed-in view of the trajectories very close to the wall.}
\label{fig:bc_test_trajectories}
\end{figure}

In this subsection, we carefully study how different implementations of the reflecting walls
(see Sec.\ \ref{subsec:wall_def}) affect the probability density of FBM on a finite interval.
Figure \ref{fig:bc_test_trajectories} presents example trajectories produced by the same
noise sequence using boundary conditions (\ref{eq:FBM_recursion_elastic}), (\ref{eq:FBM_recursion_inelastic}),
(\ref{eq:FBM_recursion_stuck}), and (\ref{eq:FBM_recursion_soft}).
The figure shows that the differences between these trajectories are of the order of the step size $\sigma$
while they become indistinguishable on length scales large compared to $\sigma$.

\begin{figure}
\includegraphics[width=\columnwidth]{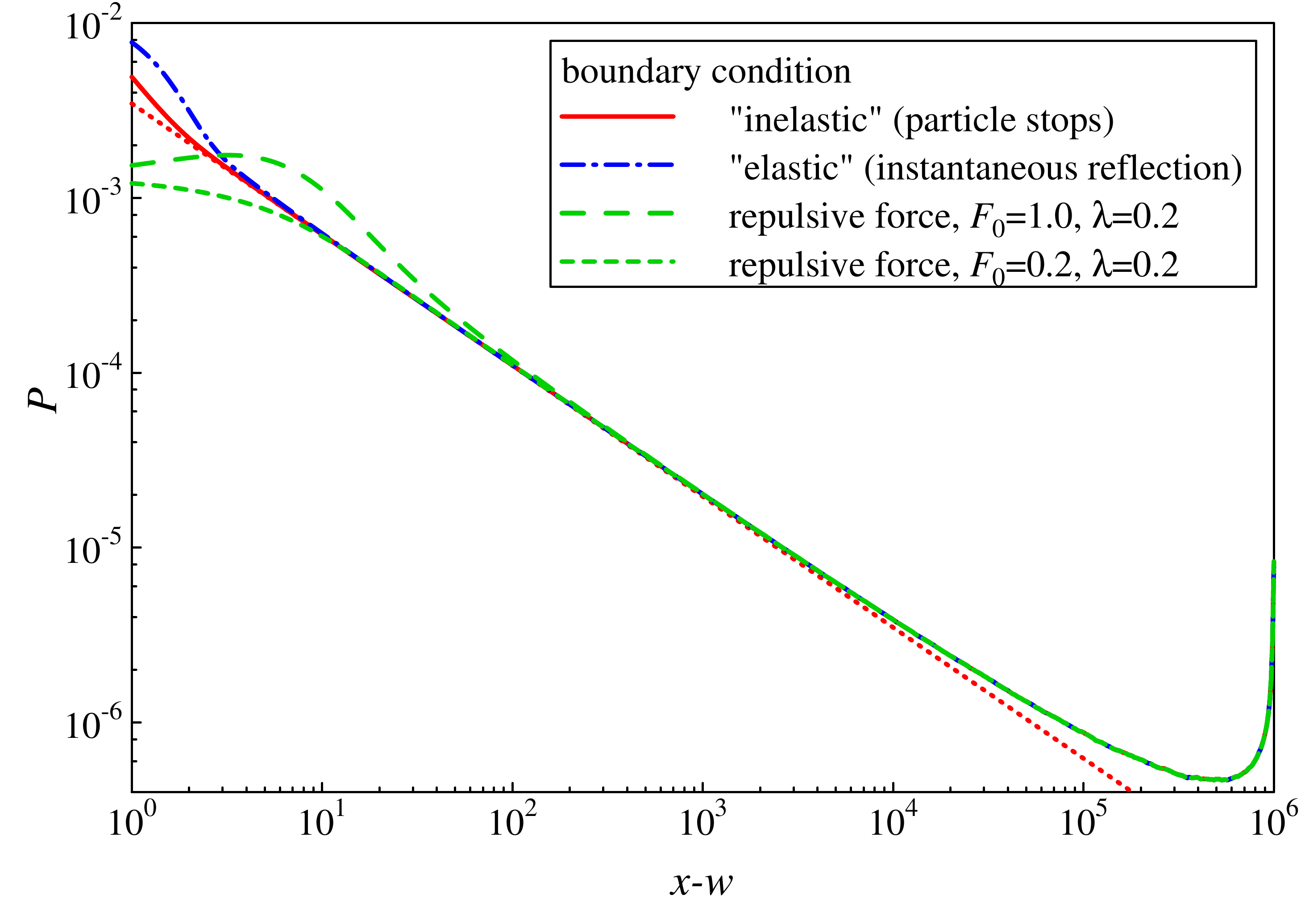}
\caption{Stationary probability density $P$ vs.\ distance $x-w$ from the wall for $\alpha=1.6$,
$L=10^6$ and several implementations of the reflecting walls. Each distribution is based on
$10^5$ particles; $P$ is averaged over a large number of time steps in the stationary regime.
The dotted line is a power law fit.}
\label{fig:1d_PDF_wall_comparison}
\end{figure}

To analyze the effects of the wall implementation quantitatively,
we compare in Fig.\ \ref{fig:1d_PDF_wall_comparison} the stationary probability densities $P(x)$
for ``elastic'' walls (\ref{eq:FBM_recursion_elastic}), ``inelastic'' walls (\ref{eq:FBM_recursion_inelastic}), and
walls implemented via ``soft'' repulsive forces (\ref{eq:wall_force}) with two different
amplitudes for a finite interval of length $L=10^6$ and $\alpha=1.6$.
The data show that the wall implementation indeed influences the probability density
in the immediate vicinity of wall. For example, a strong repulsive force pushes the peak
of $P(x)$ away from the nominal wall position. However, the figure also demonstrates that
all four wall implementations produce exactly the same probability density further away
from the wall.

\begin{figure}
\includegraphics[width=\columnwidth]{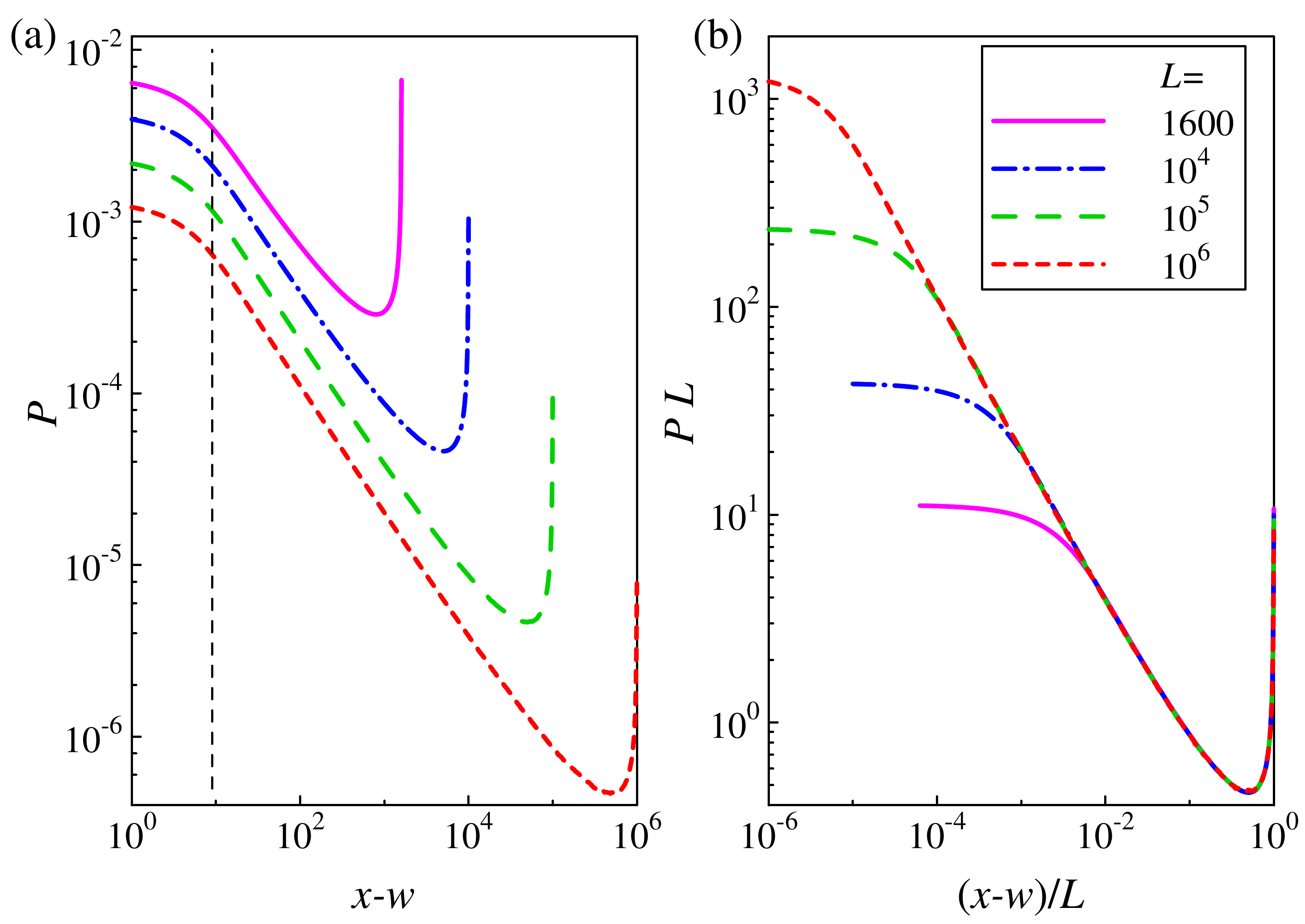}
\caption{(a) Stationary probability density $P$ vs.\ distance $x-w$ from the wall for $\alpha=1.6$,
and several interval lengths. The walls are implemented as repulsive forces (\ref{eq:wall_force})
with $F_0=0.2$ and $\lambda=0.2$. The dashed line marks the widths of the wall region.
(b) The same data plotted in scaled variables $P L$ vs.\ $(x-w)/L$.}
\label{fig:1d_PDF_wall_comparison_all_L}
\end{figure}

To investigate in more detail the region in which the wall implementation affects
$P(x)$, we compare the results for different interval lengths. Figure
\ref{fig:1d_PDF_wall_comparison_all_L} presents the stationary probability density
for $\alpha=1.6$ for intervals having lengths from $L=1600$ to $10^6$ employing
``soft'' walls defined by the repulsive force (\ref{eq:wall_force}) with amplitude
$F_0=0.2$ and decay constant $\lambda=0.2$.
The left panel, Fig.\ \ref{fig:1d_PDF_wall_comparison_all_L}a, indicates that the
width of the wall region (marked by the dashed line) is independent of the interval
lengths. This implies that the wall region becomes unimportant for
$L \gg \sigma, L \gg \lambda^{-1}$. Indeed, the right panel, Fig.\
\ref{fig:1d_PDF_wall_comparison_all_L}b, shows that the same data, plotted in scaled variables
$P L$ vs.\ $(x-w)/L$, collapse onto a common master curve for $x$ outside of each of the respective
wall regions.

These results demonstrate that variations of the probability density due to different
implementations of the reflecting walls can be considered finite-size effects that vanish
in the continuum limit $L/\sigma \to \infty, L\lambda \to \infty$. A rigorous proof
that the discretization error vanishes in the continuum limit was given in Ref.\
\cite{McGlaughlinChronopoulou16} for the wall implementation (\ref{eq:FBM_recursion_stuck}).

Note that inside the wall region (distances of order $\sigma$ from the wall), some implementations
of the reflecting boundary are better behaved than others and converge faster to the continuum limit,
as was shown for normal diffusion in Refs.\ \cite{SzymczakLadd03,NandigamKroll07}.
For example, we observed in Ref.\ \cite{Guggenbergeretal19} that the wall implementation
(\ref{eq:FBM_recursion_stuck}) leads to stronger discretization artifacts than rule (\ref{eq:FBM_recursion_elastic}).
However, all of these artifacts vanish in the continuum limit.

This differs from the behavior of the fractional Langevin equation
with reflecting walls where recent computer simulations
\cite{VojtaSkinnerMetzler19} have shown that the implementation
of the wall appears to affect the probability density in the entire interval,
perhaps due to a subtle interplay of the boundary conditions and the fluctuation-dissipation
theorem that establishes thermal equilibrium.

\subsection{FBM in superharmonic potentials}
\label{subsec:superharmonic}

In this subsection, we briefly address the behavior of FBM that is confined to a finite interval
not by reflecting walls but by a smooth external potential. The goal is to further underline
that the observed accumulation and depletion effects are neither artifacts of the specific
reflecting boundary conditions considered in the remainder of this paper, nor due to the implementation
of the fractional Gaussian noise (the noise sequence, once simulated, is used as
input continuously, no matter whether a reflection takes place or not).
For a harmonic
potential $U(x)\propto x^2$, FBM can be solved exactly and was analyzed in detail
in Refs.\ \cite{Sliusarenko10,JeonMetzler10}. In particular, the probability density remains
Gaussian in this case. However, if we consider somewhat steeper
potentials, for instance, the quartic form $U(x)\propto x^4$, distinct deviations
from the naively expected Boltzmann form $P(x)\propto\exp(-a x^4)$ can be observed.
In this case, the time evolution of the process can be obtained from the discrete
Langevin equation (\ref{eq:FBM_recursion_soft}) with $F(x)=-dU/dx$.

\begin{figure}
\includegraphics[width=\columnwidth]{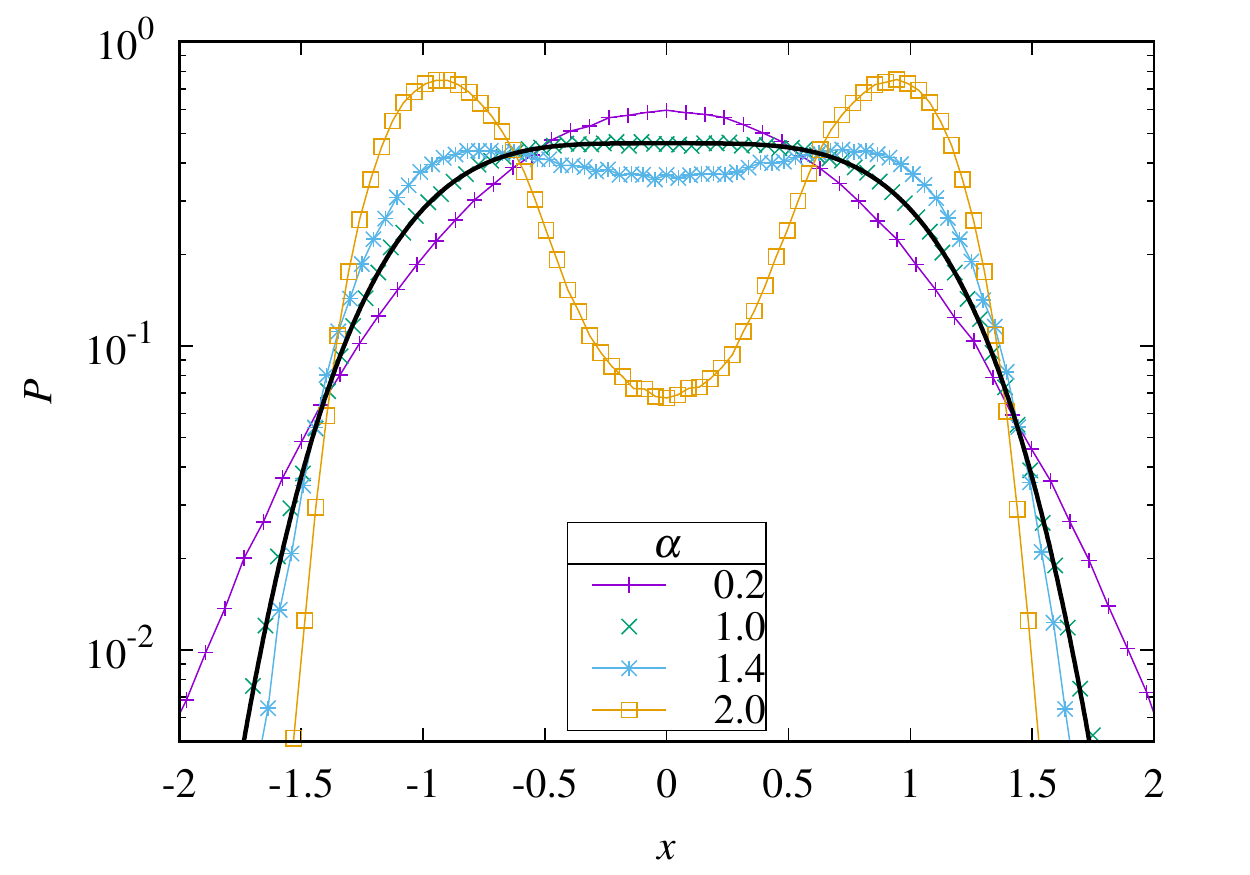}
\caption{Stationary probability density $P$ of FBM in a quartic potential
$U(x)\propto x^4$ for several values of $\alpha$. Each distribution is based on
5000 time steps, averaged over 500\,000 trajectories.
For normal Brownian motion, $\alpha=1$,
the Boltzmann distribution (heavy black line) fits the simulation results very well,
whereas in the sub- and superdiffusive cases, respectively, depletion
and accumulation with respect to the Boltzmann law are observed close to the points
of highest curvature of $U(x)$.}
\label{fig:quartic}
\end{figure}

To study FBM in a quartic potential, we perform simulations of the recursion relation (\ref{eq:FBM_recursion_soft})
using a force $F(x_n) = -\epsilon k x_n^3$ and a noise variance of  $\sigma^2=\epsilon^\alpha$,
with $k=0.2$ and $\epsilon=0.002$.
Figure \ref{fig:quartic} shows the resulting stationary probability density for different $\alpha$.
In the case of normal Brownian motion, $\alpha=1$, the Boltzmann form is reproduced
very well. However, relative to the Boltzmann law, the probability density near
the points of highest curvature of the external potential is increased for
superdiffusive FBM ($\alpha>1$) and decreased for the subdiffusive case ($\alpha<1$).
These observations are fully consistent with our
results for the reflecting boundary conditions.

\section{Two space dimensions}
\label{sec:2d}
\subsection{Overview}
\label{subsec:2d_overview}

\begin{figure}[b]
\includegraphics[width=\columnwidth]{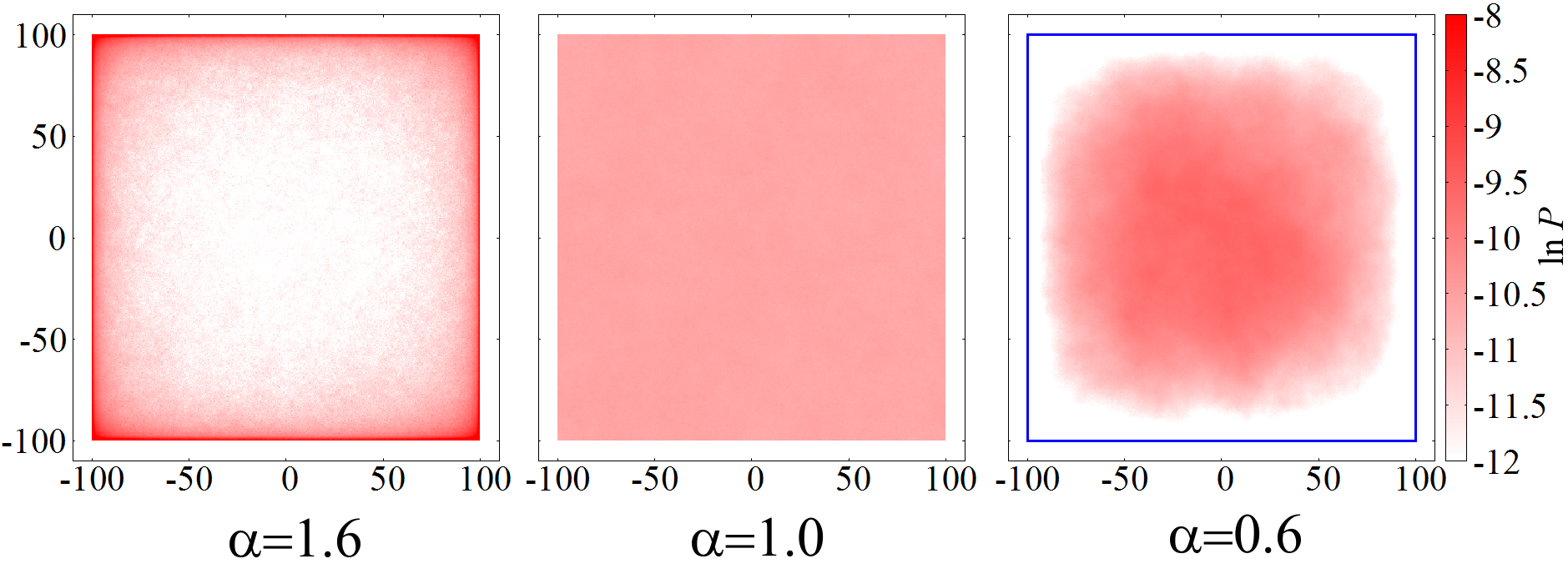}
\caption{Stationary probability density $P$ of FBM on a square domain for several $\alpha$. The heat maps
of $\ln P$ are based on 100 particles performing up to $2^{24}$ time steps each.}
\label{fig:2d_heatmaps_squares_all_alpha}
\end{figure}

Let us now turn to reflected FBM in two dimensions. We have performed simulations for a variety
of geometries. For a qualitative overview, we present in Fig.\ \ref{fig:2d_heatmaps_squares_all_alpha}
heat maps of the stationary probability density of FBM confined to a square domain by reflecting walls.
The figure compares three different values of the anomalous diffusion exponent, viz.,
$\alpha=1.6$ (superdiffusive regime), $\alpha=1$ (normal Brownian motion), and $\alpha=0.6$
(sub\-diffusive regime). The data indicate the same qualitative behavior as observed in one dimension.
In the superdiffusive regime, particles accumulate close to the reflecting boundaries, compared
to the flat distribution for normal diffusion. In the subdiffusive regime, in contrast, particles
are depleted close to the walls. The strongest accumulation and depletion are seen in the
corners of the square.

\begin{figure}
\includegraphics[width=3.6cm]{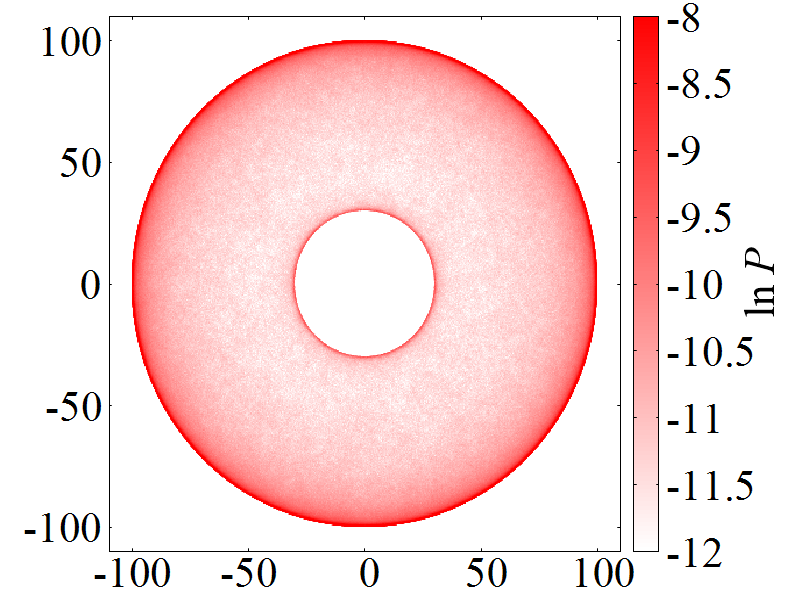}~\includegraphics[width=3.6cm]{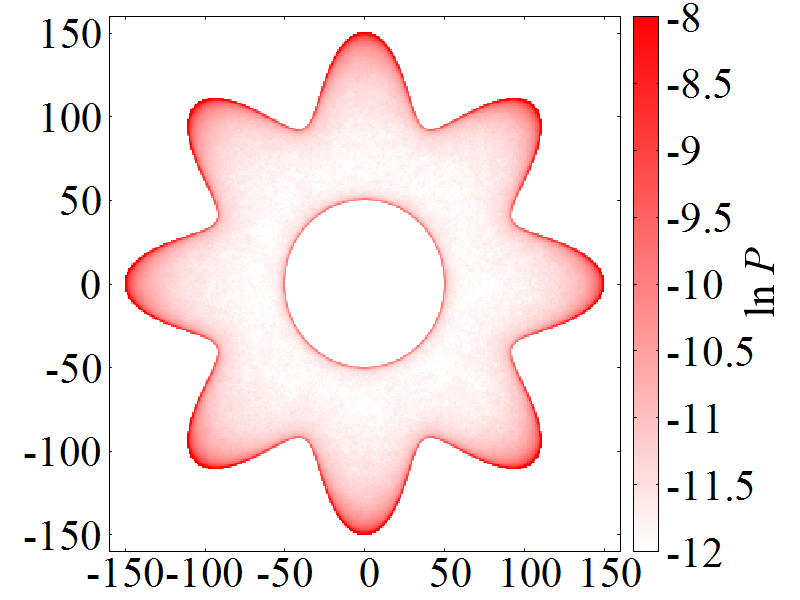}
\caption{Heat maps of the stationary probability density $P$ for $\alpha=1.6$, computed from 100 particles
performing up to $2^{21}$ time steps.}
\label{fig:2d_heatmaps}
\end{figure}

Analogous accumulation and depletion effects are also observed in other geometries.
Figure \ref{fig:2d_heatmaps} shows heat maps of the stationary probability density of FBM on
a ring-shaped domain and a star-shaped domain for $\alpha =1.6$ in the superdiffusive regime.
These shapes allow us to analyze the differences between concave and convex boundaries.
As above, the data indicate that particles accumulate close to all reflecting walls. The accumulation
is stronger for concave boundaries such as the outer boundary of the ring and weaker for convex boundaries
such as its inner boundary.

\subsection{Rectangular domains}
\label{subsec:rectangle}

We now analyze the probability density of reflected FBM in two-dimensional geometries quantitatively.
Square and rectangular domains are particularly simple cases because the motions parallel and perpendicular
to the walls, i.e., the $x$ and $y$ components of the two-dimensional FBM for appropriately chosen
coordinate axes, completely decouple \footnote{$x$ and $y$ may be coupled during the reflection process
for some choices of the reflection condition. Based on the results of Sec.\ \ref{subsec:wall_test}, this is
not expected to influence the probability density outside the narrow ``wall region''.}.
The two-dimensional probability density is therefore simply a product of two one-dimensional probability
density functions. Specifically, for a rectangle of sides $L_x$ and $L_y$, the stationary probability
density takes the form
\begin{equation}
P_{2d}(x,L_x;y,L_y) = P_{1d}(x,L_x) P_{1d}(y,L_y)
\label{eq:P_rectangle}
\end{equation}
in the continuum (scaling) limit $L_x, L_y \gg \sigma$.  Here, $P_{1d}(x,L_x)$ and $P_{1d}(y,L_y)$ are the
stationary distributions of one-dimensional FBM on finite intervals of length $L_x$ and $L_y$, respectively.

This has the following implications for behavior of the stationary probability density at the boundaries of the
rectangular domain. When the edge of the rectangle is approached away from a corner, the probability density
features a power-law singularity with the same exponent value, $\kappa=2/\alpha-2$, as in one dimension. In
contrast, if the corner of the rectangle is approached along the diagonal (or any other straight line), the
probability density follows a power-law with the doubled exponent $\kappa=4/\alpha-4$.
Consistently, relatively higher densities are observed close to the corners.
We have confirmed this explicitly by computer simulations on square domains for anomalous diffusion exponents
$\alpha=0.8, 1.2, 1.4$ and 1.6.

\subsection{Disks and rings}
\label{subsec:disks}

For FBM on domains with curved boundaries, such as a circular domain (disk) of
radius $R$, the situation is more complicated.
For uncorrelated or short-range correlated random walks, one would expect the
curvature of the boundary to become unimportant if the radius $R$ of the curvature
is large compared to the step size $\sigma$ (or the finite correlation length of the steps).
However, FBM has long-range correlations and thus effectively sees (remembers)
the entire domain. It is therefore not clear a priori whether or not the curvature
affects the behavior of the probability density near the boundary.

\begin{figure}
\includegraphics[width=\columnwidth]{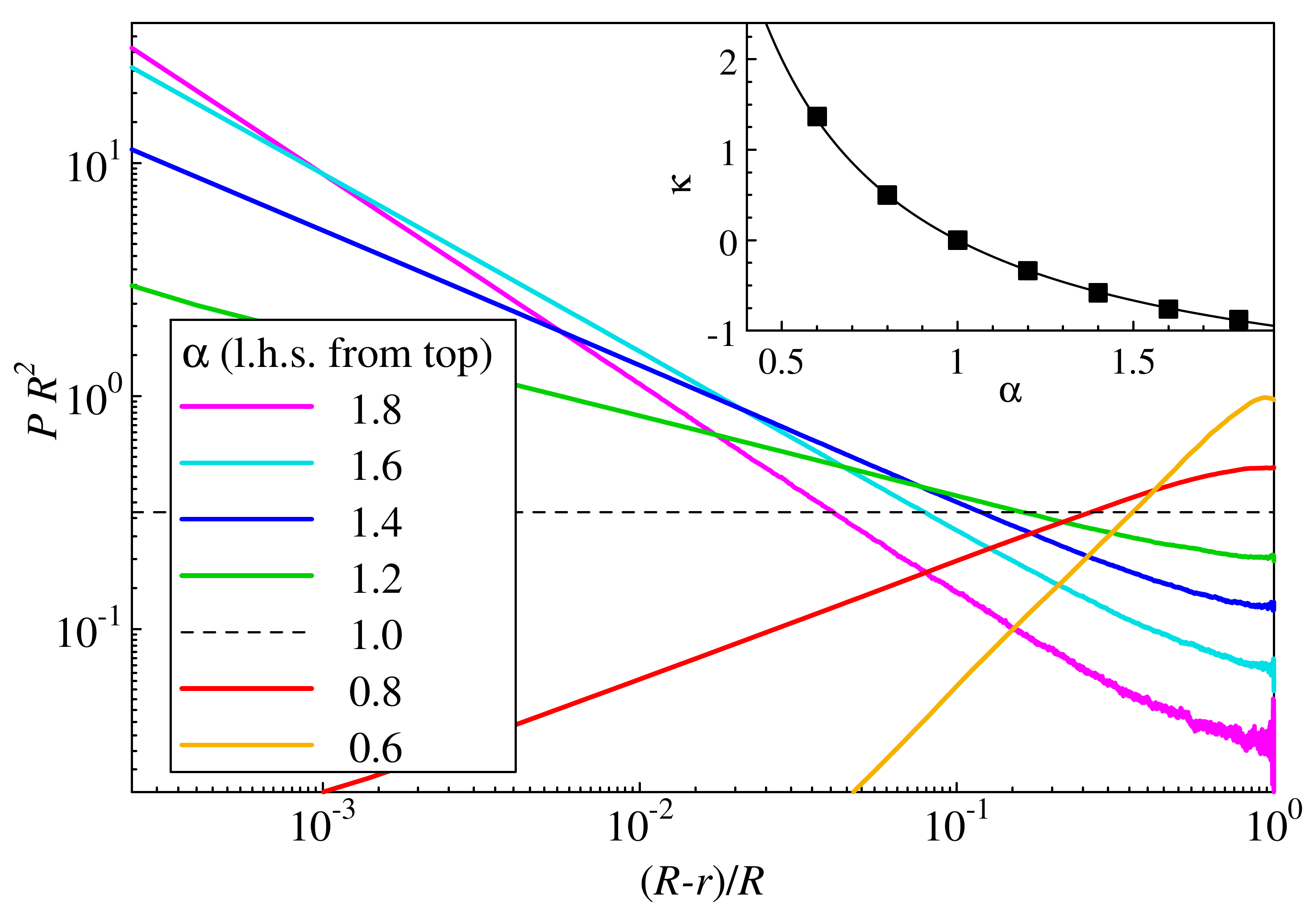}
\caption{Stationary probability density $P(r)$ of FBM on circular domains (disks) of radius $R$ for several $\alpha$.
The data are plotted as $P R^2$ vs.\ scaled distance $(R-r)/R$ from the wall. System sizes range
from $R=10^6$ for $\alpha=1.8$ to $R=1000$ for $\alpha=0.6$; the simulations use $10^4$ to $10^5$
particles and up to $2^{29}$ time steps.
Inset: Exponent $\kappa$ extracted from power-law fits of $P(r)$ close to the wall. The solid line
is the one-dimensional conjecture $\kappa =2/\alpha -2$. }
\label{fig:PDF_log_scaled_all_disk}
\end{figure}

To resolve this question, we perform extensive simulations of FBM on large circular domains with radii up to
$R=10^6$ for anomalous diffusion exponents $\alpha$ between 0.6 and 1.8. We find that the stationary
probability density is, of course,  rotationally invariant, i.e., independent of the polar angle.
Its radial dependence fulfills the scaling form
\begin{equation}
P_{2d}(r,R) = \frac 1 {R^2} Y_\alpha(r/R)
\label{eq:P_scaling_R}
\end{equation}
for $R \gg \sigma$. Here, $r$ is the distance from the center of the disk.
Figure \ref{fig:PDF_log_scaled_all_disk} summarizes the results of these simulations,
focusing on the behavior of the probability density $P$ close to the reflecting boundary at
$r=R$.
It shows that $P$ behaves as a power of the distance from the wall for all $\alpha$. We determine the
exponent from fits of the power law $P(r) \sim (R-r)^\kappa$ to the probability density close to the
wall but outside of the region influenced by finite-size effects, i.e., for $\sigma \ll R-r \ll R$.
The inset of Fig.\ \ref{fig:PDF_log_scaled_all_disk} shows the resulting values of the exponent $\kappa$
as a function of $\alpha$. They follow the same conjecture $\kappa=2/\alpha -2$ as in the one-dimensional
case, suggesting that the curvature of the reflecting wall does not affect the functional form of the
probability density near the wall.

\begin{figure}
\includegraphics[width=\columnwidth]{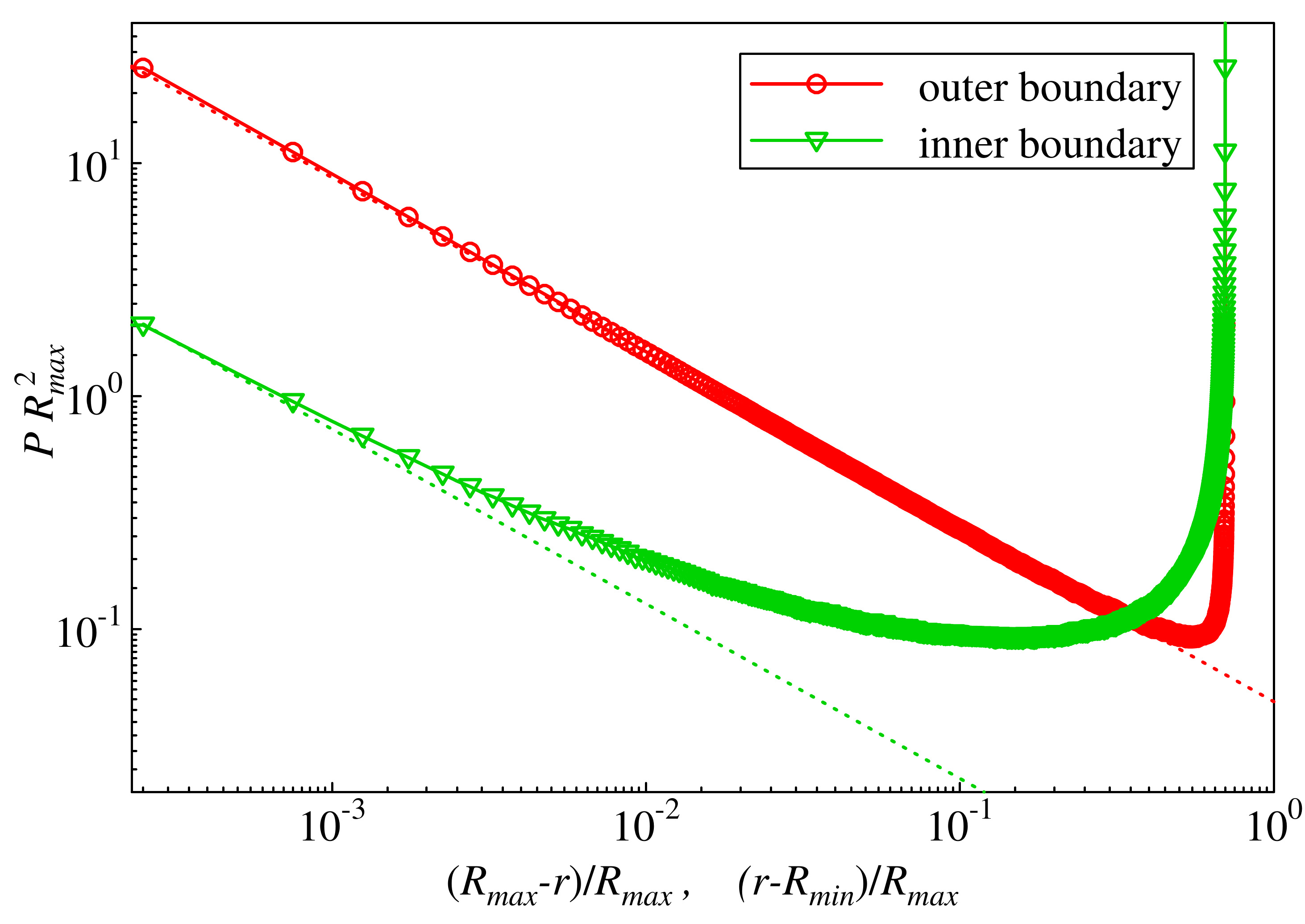}
\caption{Scaled stationary probability density on a ring with outer radius $R_{max}=10^6$ and inner radius
$R_{min}=0.3 R_{max}$ vs.\ scaled distance from both boundaries for $\alpha =1.6$. The simulations use
$10^5$ particles performing $2^{27}$ time steps. The dashed lines are fits to power laws with the
conjectured exponent $\kappa=2/\alpha -2 = -0.75$. }
\label{fig:PDF_log_scaled_ring}
\end{figure}

In addition to disks, we also consider ring-shape domains. As was already shown in the heat map in Fig.\
\ref{fig:2d_heatmaps}, particles accumulate at both the inner and the outer boundary of the ring for
superdiffusive FBM. However, the accumulation is stronger at the concave outer boundary than at the
convex inner boundary. Figure \ref{fig:PDF_log_scaled_ring} presents a quantitative analysis of the
probability density close to both walls for $\alpha=1.6$.
Close to the outer boundary, the probability density clearly follows the conjectured power law
$P \sim (R_\textrm{max} -r)^{2/\alpha -2}$. At the inner boundary we observe a much slower crossover,
but the data are compatible with an asymptotic power-law singularity with the same exponent,  $P \sim (r -R_\textrm{min})^{2/\alpha -2}$.

\subsection{Circular sectors}

The results of the last subsection show that the curvature of a reflecting wall does not influence the
qualitative behavior of the probability density close to the wall. However, the example of a square domain
in Sec.\ \ref{subsec:2d_overview} indicates that sharp corners lead to stronger singularities of the
probability density at the boundary.

\begin{figure}
\includegraphics[height=3.6cm]{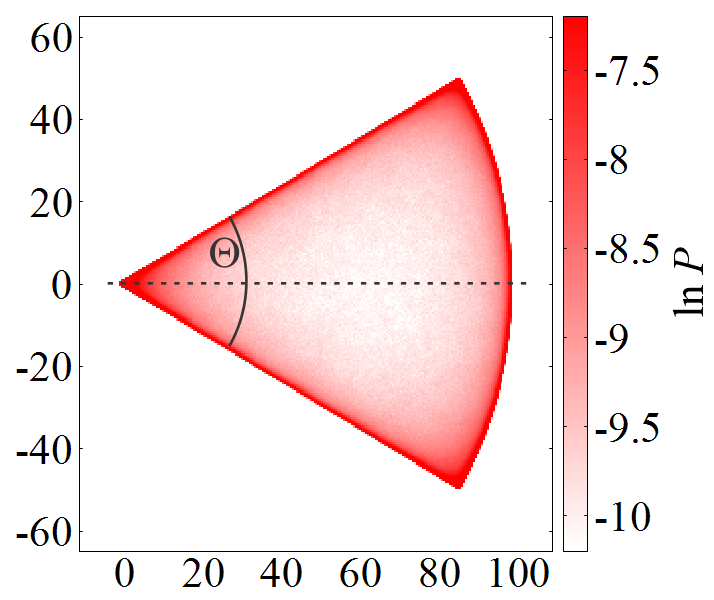}~\includegraphics[height=3.6cm]{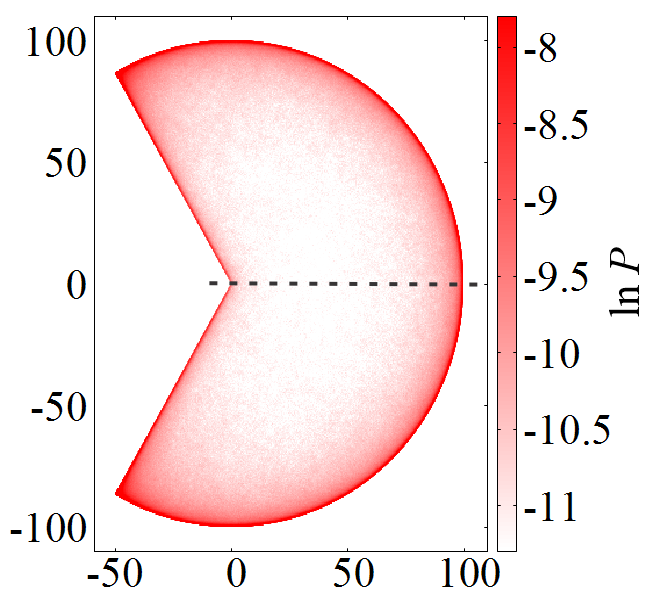}
\caption{Heat maps of the stationary probability density of FBM ($\alpha=1.6$) on circular sectors with
opening angles $\Theta=60^\circ$ and $240^\circ$. The simulations use 100 particles performing $2^{18}$ time steps.
The dotted line marks the cut used to analyze the singularity of the PDF in the tip (corner) at the center
of the curvature. }
\label{fig:sector_heatmaps}
\end{figure}

In the present section, we study this effect systematically by performing
simulations of FBM on circular sectors of radius $R$ and varying opening angle $\Theta$
for anomalous diffusion exponents $\alpha=1.6$ (superdiffusive regime) and
0.8 (subdiffusive regime).
Two examples that illustrate the geometry of these sectors are presented
in Fig.\ \ref{fig:sector_heatmaps}.
For $\alpha=1.6$, the heat map of the probability density of the $60^\circ$ sector in Fig.\ \ref{fig:sector_heatmaps} shows a particularly strong accumulation in the tip (center of curvature) of the sector.

\begin{figure}
\includegraphics[width=\columnwidth]{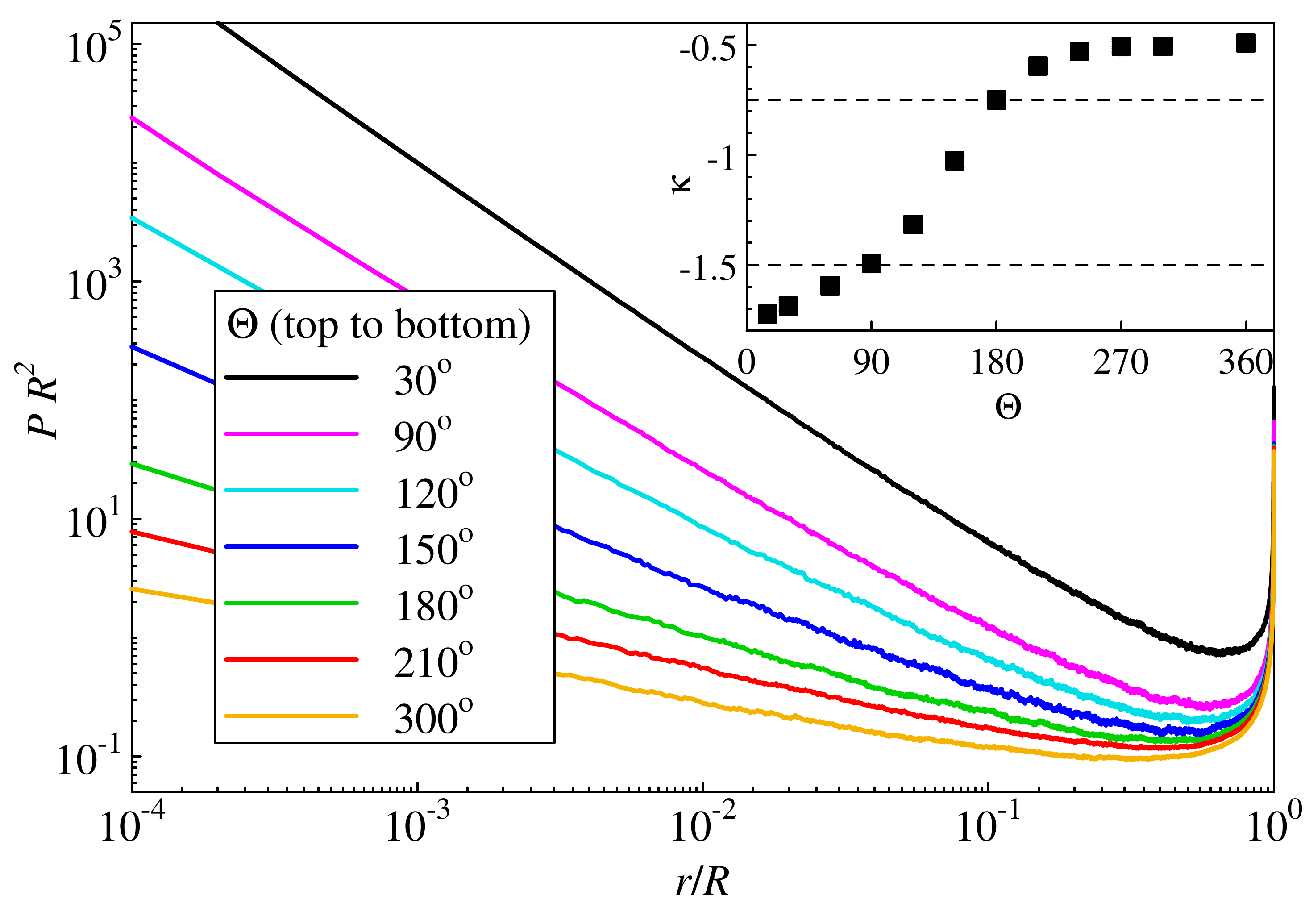}
\caption{Scaled stationary probability density $P R^2$ of FBM with $\alpha=1.6$ on circular sectors
with outer radius $R=10^5$ and various opening angles $\Theta$. The graph shows $P R^2$ on the symmetry
line of the sector as a function of the scaled distance $r/R$ from the center of curvature.
($10^5$ to $10^6$ particles performing $2^{23}$ time steps.) Inset:
Exponent $\kappa$, extracted from power-law fits,  $P(r) \sim r^\kappa$, of the data close
to the center ($r/R \ll 1$) vs.\ opening angle $\Theta$.
The dashed lines mark the values $2/\alpha-2=-0.75$ and $4/\alpha-4=-1.5$. }
\label{fig:PDF_log_scaled_sector_gamma04}
\end{figure}

To understand the behavior in the tip quantitatively, we analyze the stationary probability density
along the symmetry line (dashed line in Fig.\ \ref{fig:sector_heatmaps}) of the sector.
Figure \ref{fig:PDF_log_scaled_sector_gamma04} shows a double logarithmic plot of the (scaled)
probability density as a function of the distance from the tip.
All curves feature power law behavior for $r \ll R$ but the exponent changes continuously with the
opening angle $\Theta$ of the sector. The inset of Fig.\ \ref{fig:PDF_log_scaled_sector_gamma04}
presents the values of the exponent, determined from fits of the probability density by
$P(r) \sim r^\kappa$ for $\sigma \ll r \ll R$. We observe that the divergence of $P(r)$ becomes stronger
($\kappa$ becomes more negative) as the opening of the sector gets narrower.
For $\Theta=180^\circ$, $\kappa$ takes the value $2/\alpha-2$ as on a one-dimensional interval.
This is expected because the left boundary of the sector is a straight line for $\Theta=180^\circ$.
Similarly, we find $\kappa = 4/\alpha -4$ for $\Theta=90^\circ$, as in the corner of a square.
For $\Theta \to 360^\circ$, the exponent $\kappa$ approaches a value close to $-0.5$. At first glance, one
might have expected $\kappa$ to approach zero in this limit because the probability density of a disk
does not have a singularity in the center. Note, however, that a reflecting line along the negative $x$
axis remains in the $\Theta\to 360^\circ$ limit of the sector.

\begin{figure}
\includegraphics[width=\columnwidth]{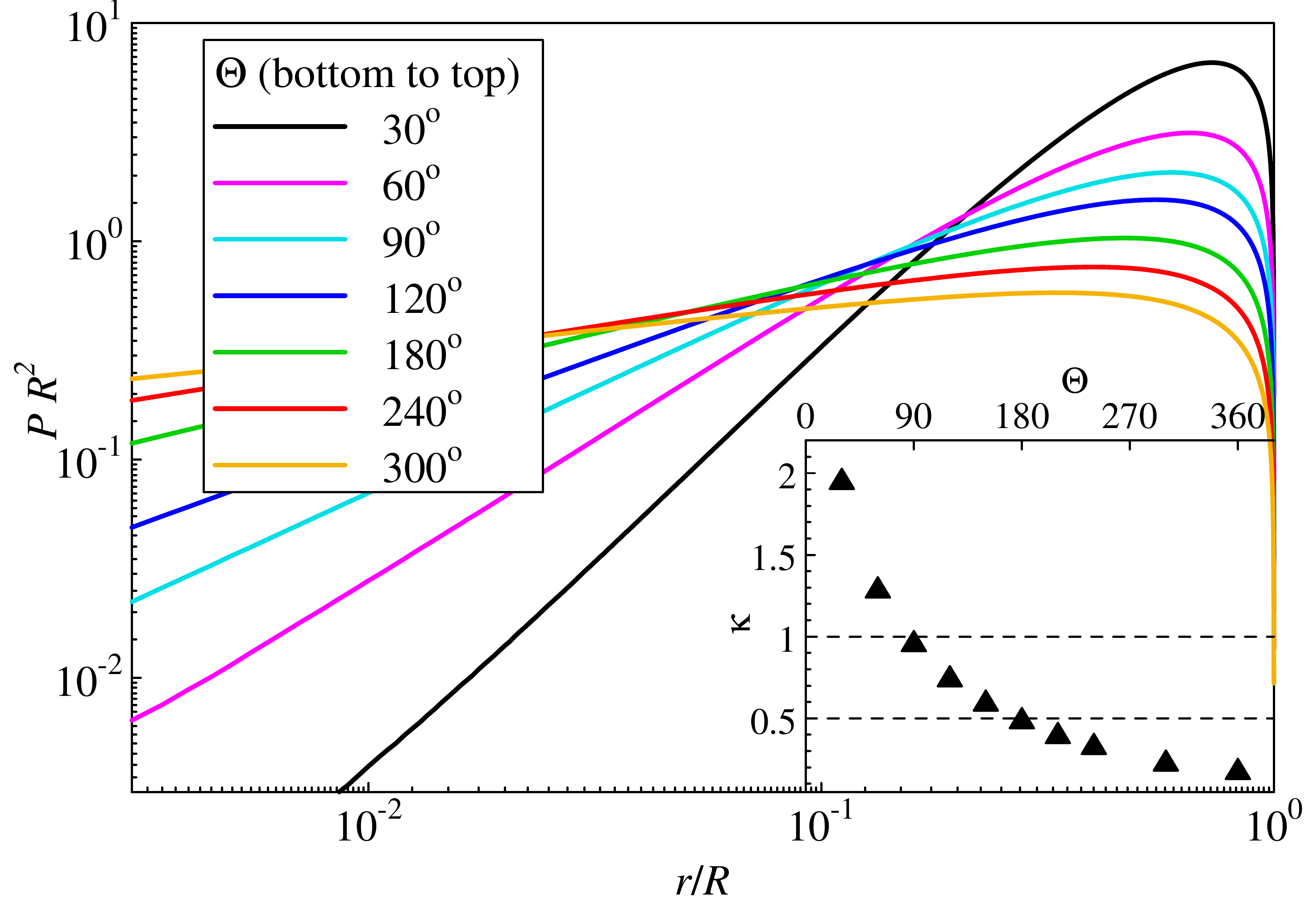}
\caption{Scaled stationary probability density $P R^2$ of FBM with $\alpha=0.8$ on a circular sector with outer radius $R=10^3$ for various opening angles
$\Theta$. The graphs show $P R^2$ on the symmetry line of the sector as a function of the scaled distance $r/R$ from the center of curvature. ($10^6$ particles performing $2^{25}$ time steps.) Inset:
Exponent $\kappa$, extracted from power-law fits,  $P(r) \sim r^\kappa$, of the PDF close to the center ($r/R \ll 1$) vs.\ opening
angle $\Theta$. The dashed lines mark the values $2/\alpha-2=0.5$ and $4/\alpha-4=1.0$. }
\label{fig:PDF_log_scaled_sector_gamma12}
\end{figure}

We also carry out analogous simulations for subdiffusive FBM using $\alpha=0.8$.
The results are presented in Fig.\ \ref{fig:PDF_log_scaled_sector_gamma12}.
As above, the deviations from a flat distribution become stronger as the opening angle of the
sector decreases. For $\Theta=180^\circ$ and $\Theta=90^\circ$, we recover the expected exponent
values $\kappa=2/\alpha-2$ and $4/\alpha-4$, respectively.

The results in this section are obtained using the ``inelastic'' boundary conditions
(\ref{eq:FBM_recursion_inelastic_d}). To confirm that the details of the wall implementation
do not affect the results, we also perform simulations using soft walls, defined by
appropriate generalizations of Eqs.\ (\ref{eq:FBM_recursion_soft}) and (\ref{eq:wall_force})
to the circular sector geometry. Specifically, we have analyzed sectors with
openings of $15^\circ$ and $90^\circ$ for $\alpha=1.6$ in this way.
As in one dimension (Sec.\ \ref {subsec:wall_test}),
we find that  the wall implementation only influences a narrow ``interaction region''
close to the wall that becomes unimportant in the continuum (scaling) limit $R/\sigma \to \infty$.

\section{Three space dimensions}
\label{sec:3d}

In this section, we briefly discuss reflected FBM in three dimensional geometries.
Domains shaped as rectangular prisms (cuboids) can be analyzed analogously to Sec.\
\ref{subsec:rectangle}. Because the $x$, $y$, and $z$ components of a three-dimensional
FBM are independent of each other, the stationary probability density of FBM in a
rectangular prism of sides $L_x$, $L_y$, and $L_z$ factorizes and takes the form
\begin{equation}
P_{3d}(x,L_x;y,L_y;z,L_z) = P_{1d}(x,L_x) P_{1d}(y,L_y) P_{1d}(z,L_z)
\label{eq:P_prism}
\end{equation}
in an appropriate coordinate system having axes parallel to the edges of the prism.
This implies that the probability density features a power-law singularity with
exponent $2/\alpha -2$ when a face of the prism is approached. If an edge is approached
the exponent is given by $4/\alpha -4$, and when a corner is approached (along a straight line)
the exponent is expected to be $6/\alpha -6$.

Turning to spherical domains, we simulate superdiffusive FBM with $\alpha=1.6$ in a sphere
of radius $R=10^6$ and subdiffusive FBM with $\alpha=0.8$ in a sphere of radius $R=10^3$.
We observe that the behavior of the stationary probability density is completely analogous
to the case of a circular (disk) domain discussed in Sec.\ \ref{subsec:disks}. Specifically,
the probability density features a power-law singularity at the surface of the sphere
that is controlled by the one-dimensional exponent $\kappa=2/\alpha -2$.

\begin{figure}
\includegraphics[width=4.3cm]{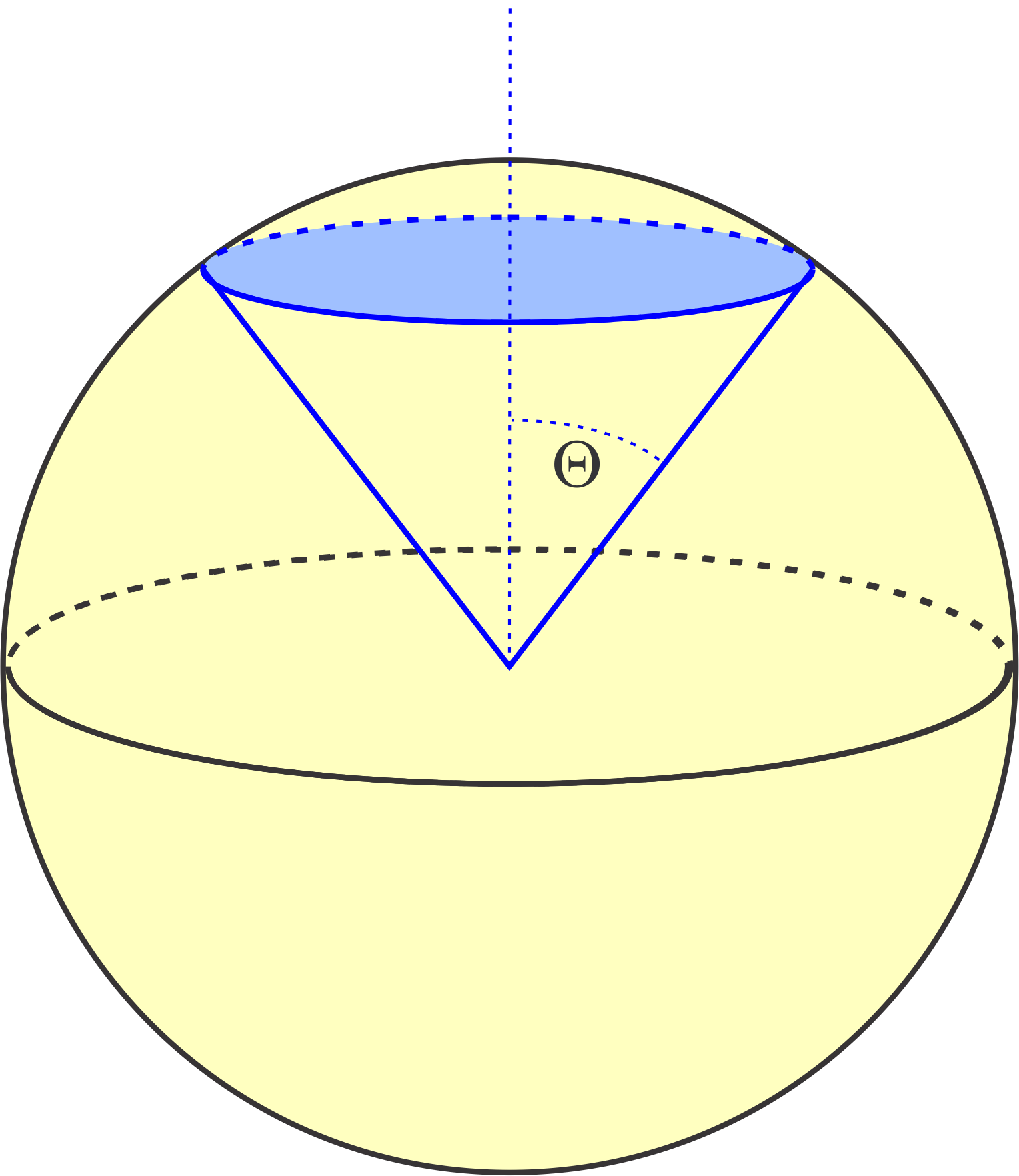}
\caption{Geometry of the spherical cone. }
\label{fig:cone}
\end{figure}

\begin{figure}
\includegraphics[width=\columnwidth]{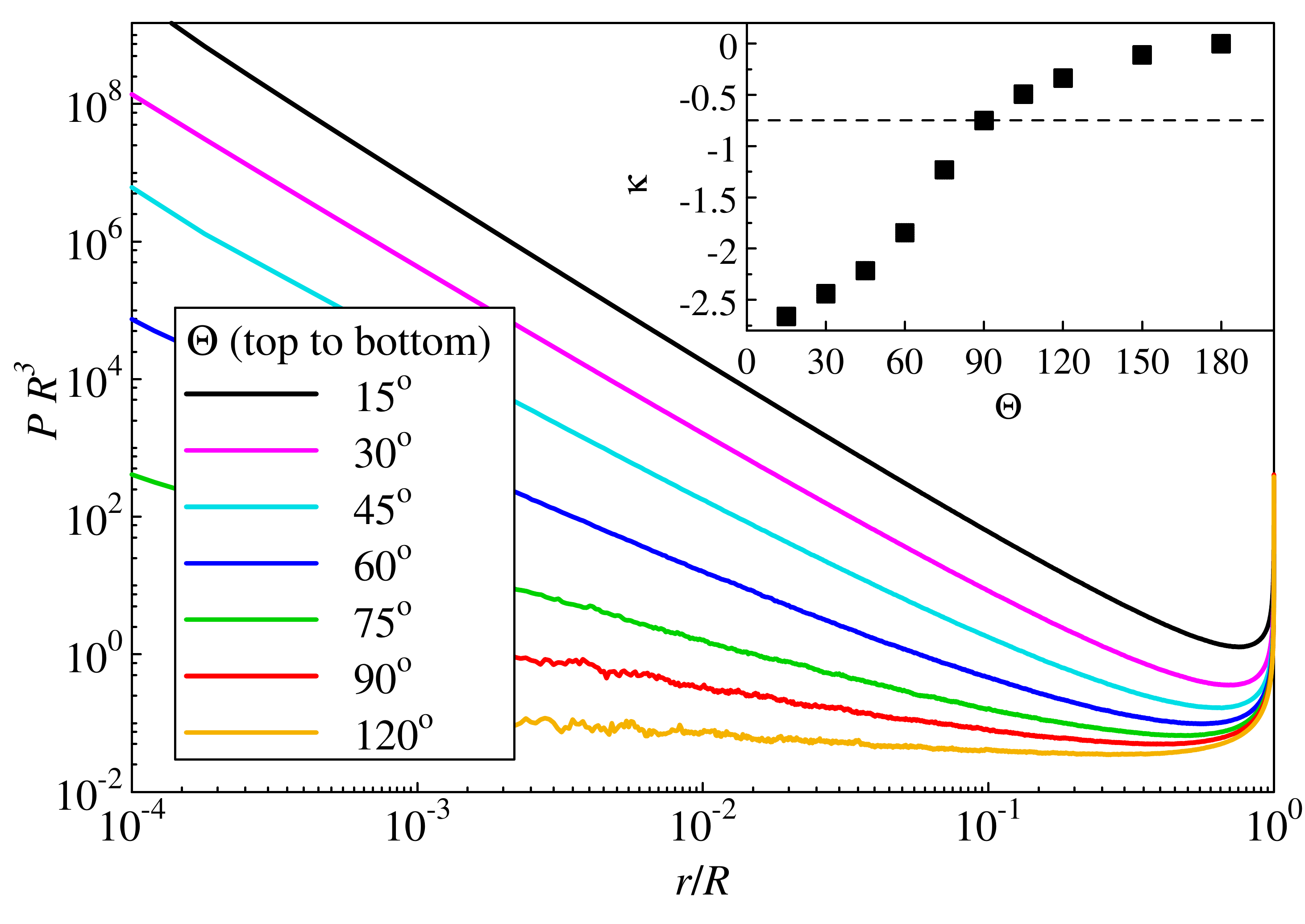}
\caption{Scaled stationary probability density $P R^3$ of FBM with $\alpha=1.6$ in a spherical cone with outer radius $R=10^5$ for various opening angles $\Theta$.  The graphs show $P R^3$ on the symmetry line of the cone
vs.\ the scaled distance $r/R$ from the center of curvature. ($10^7$ particles performing $2^{23}$ time steps.) Inset:
Exponent $\kappa$, determined from power-law fits,  $P(r) \sim r^\kappa$, of the data close to the center ($r/R \ll 1$) vs.\ opening angle $\Theta$. The dashed line marks the value $2/\alpha-2=-0.75$. }
\label{fig:PDF_log_scaled_cone_gamma04}
\end{figure}

To determine how the ``sharpness'' of a corner affects the probability density in three dimensions, we
simulate FBM in spherical sectors (spherical cones) of variable opening angles for $\alpha=1.6$ and
0.8. A spherical cone contains all points whose distance from the origin is less than $R$
and whose polar angle is less than $\Theta$, see Fig.\ \ref{fig:cone}.
We then analyze the probability density on the symmetry axis of the cone.
The results for superdiffusive motion with $\alpha=1.6$ are presented in
Fig.\ \ref{fig:PDF_log_scaled_cone_gamma04} for several opening angles $\Theta$
of the cone.
As in the case of circular sectors, all curves feature power-law singularities close to
the tip (center of curvature) of the cone. We determine the exponents from fits of the
probability density to $P(r) \sim r^\kappa$. The resulting values are presented in the inset
of Fig.\ \ref{fig:PDF_log_scaled_cone_gamma04} as a function of the opening angle $\Theta$.
For $\Theta=90^\circ$, $\kappa$ takes the one-dimensional value $2\alpha-2$ because the reflecting wall
at the bottom of the cone is completely flat. For $\Theta \to 180^\circ$, the exponent
$\kappa$ approaches zero (corresponding to a nonsingular $P$) because particles can easily go around
the repulsive line along the negative $z$ axis remaining in this limit (in contrast to the
two-dimensional case).

\begin{figure}
\includegraphics[width=\columnwidth]{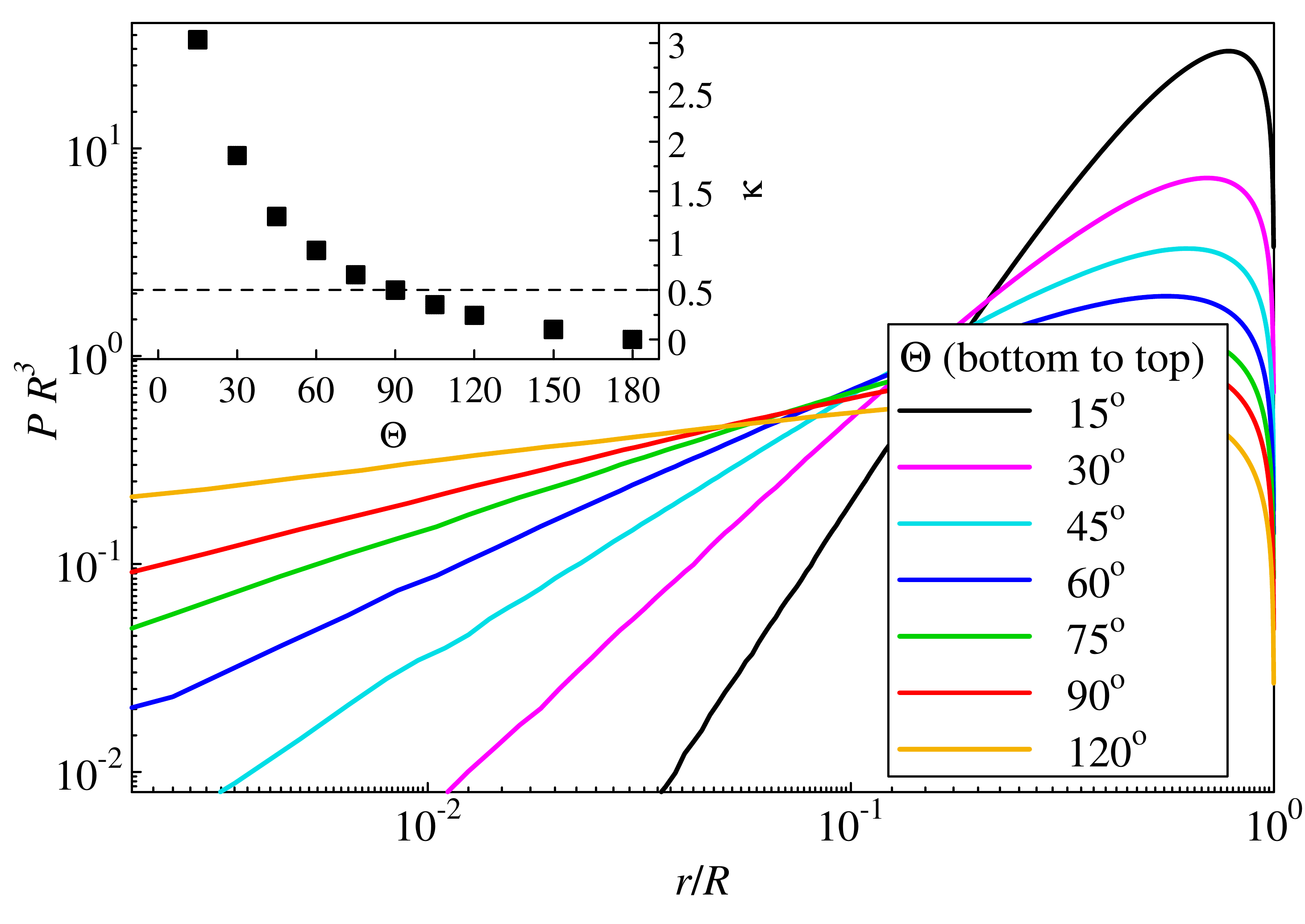}
\caption{Scaled stationary probability density $P R^3$ of FBM with $\alpha=0.8$ in a spherical cone with
outer radius $R=10^3$ for various opening angles
$\Theta$.  The graphs show $P R^3$ on the symmetry line of the cone as functions of $r/R$.
($10^6$ particles performing $2^{25}$ time steps.) Inset:
Exponent $\kappa$, determined from power-law fits,  $P(r) \sim r^\kappa$, of the data close to the
center ($r/R \ll 1$) vs.\ opening angle $\Theta$. The dashed line marks the value $2/\alpha-2=0.5$. }
\label{fig:PDF_log_scaled_cone_gamma12}
\end{figure}

Figure \ref{fig:PDF_log_scaled_cone_gamma12} presents the same analysis for subdiffusive motion
with $\alpha=0.8$.
We again observe power-law singularities in the probability density close to the tip of the cone,
with an exponent that increases continuously as the cone narrows. The values of the
scaling exponent, determined
from power-law fits, are shown in the inset of the figure. As expected, for $\Theta=90^\circ$, we
recover the one-dimensional exponent $2/\alpha-2$.

\section{Application to brain serotonergic fibers}
\label{sec:fibers}

As was pointed out in the introductory section of this paper, FBM has found a broad variety of
applications in physics, chemistry, biology, and beyond. Recently, it has been proposed that FBM
may be a good model for the geometry of serotonergic fiber paths in vertebrate brains, including the human brain.

\begin{figure}
\includegraphics[width=\columnwidth]{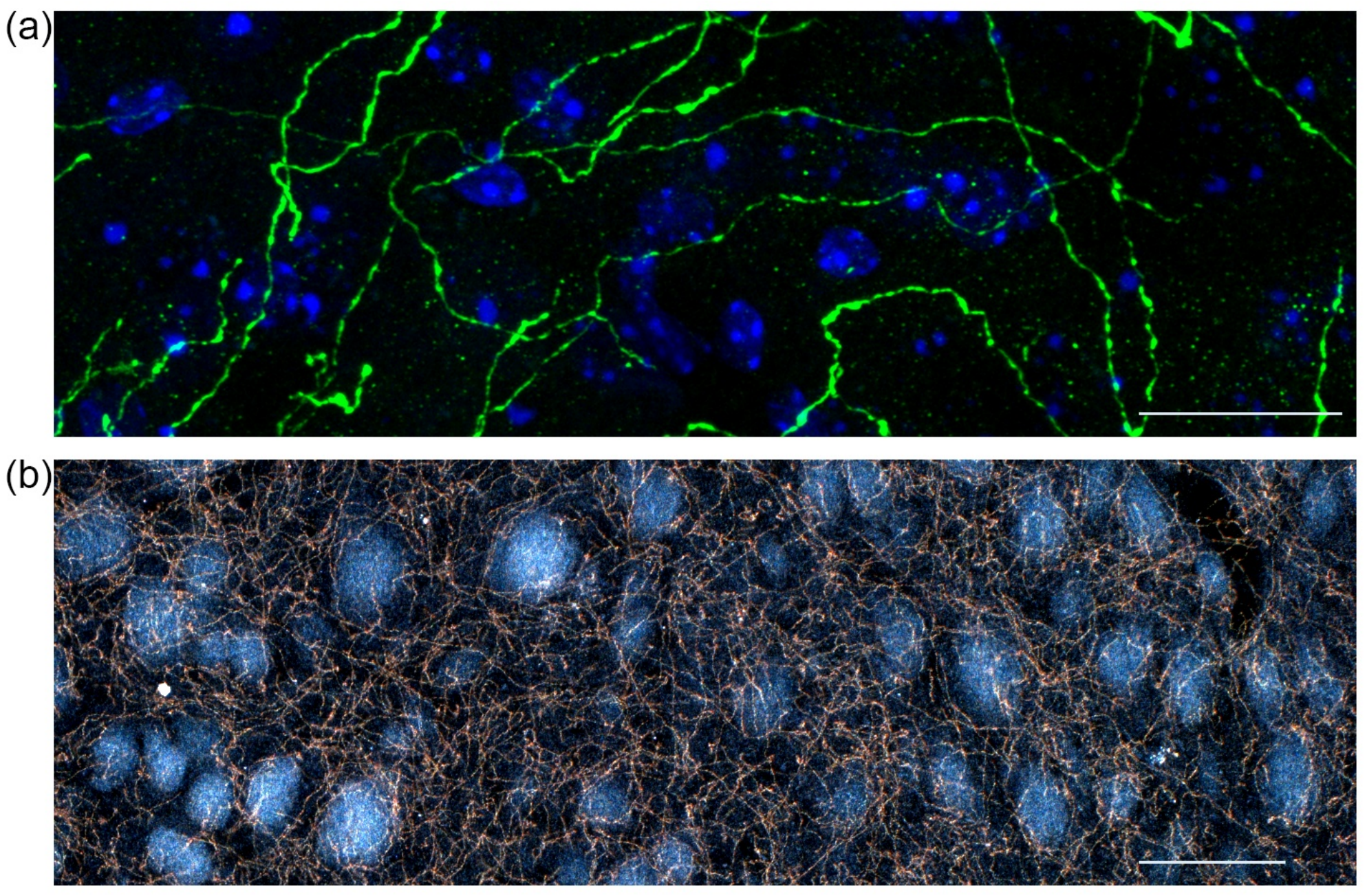}
\caption{(a) A confocal microscopy image of serotonergic fibers visualized with an anti-GFP antibody in the cingulate cortex (area 30) of the transgenic mouse model developed by Migliarini et al.\ \cite{Migliarinietal13}. The fibers 
are shown as green (bright) lines; cell nuclei are labeled blue (dark gray). Scale bar = 20 $\mu$m (the thickness 
of the z-stack is 19 $\mu$m). (b) A dark-field microscopy image of serotonergic fibers visualized with an anti-serotonin-transporter antibody in the caudate-putamen of a wild-type mouse. The fibers are shown as golden
brown (light gray) lines. Scale bar = 100 $\mu$m.}
\label{fig:real_fibers}
\end{figure}

The entire central nervous system of vertebrates is permeated by a dense network of serotonergic fibers,
very long axons of neurons that are located in the brainstem \cite{Hornung03,OkatyCommonsDymecki19}.
These fibers release the neurotransmitter
serotonin as well as other neurotransmitters. The densities of this serotonergic matrix vary
significantly across brain regions, and their perturbations  can severely affect the function of
neural circuits. Traditionally, the emergence of these densities has been treated as a tightly controlled sequence of developmental events that reflects the functional requirements of individual brain regions (neuroanatomical nuclei and laminae). Based on high-resolution imaging techniques (see Fig.\ \ref{fig:real_fibers}),
it has been suggested, however,  that individual fibers behave as
three-dimensional stochastic processes, with the varying fiber densities emerging from the
interaction of the randomness with the complex brain geometry \cite{Janusonis17,JanusonisDetering19}.
Specifically, superdiffusive FBM has emerged as a promising theoretical framework for the
description of brain serotonergic fibers \cite{JanusonisDetering19,JanusonisDeteringMetzlerVojta20}.

\begin{figure}
\includegraphics[width=4.4cm]{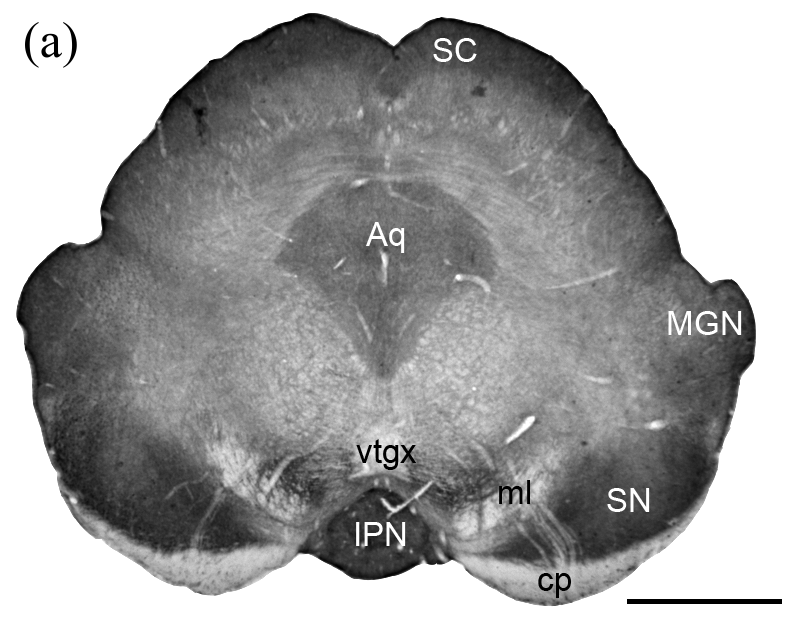}~\includegraphics[width=4.3cm]{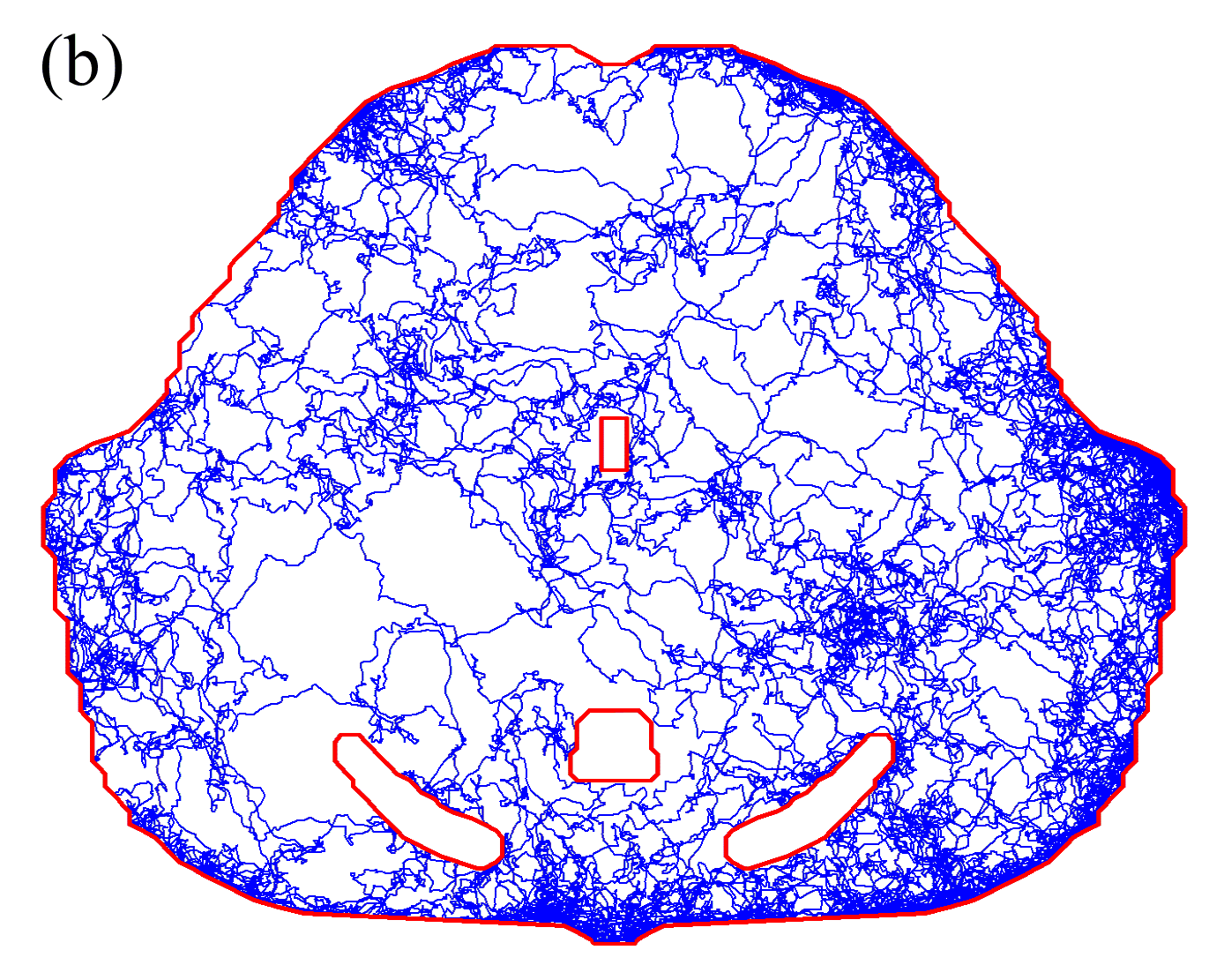} \\ \includegraphics[width=5.2cm]{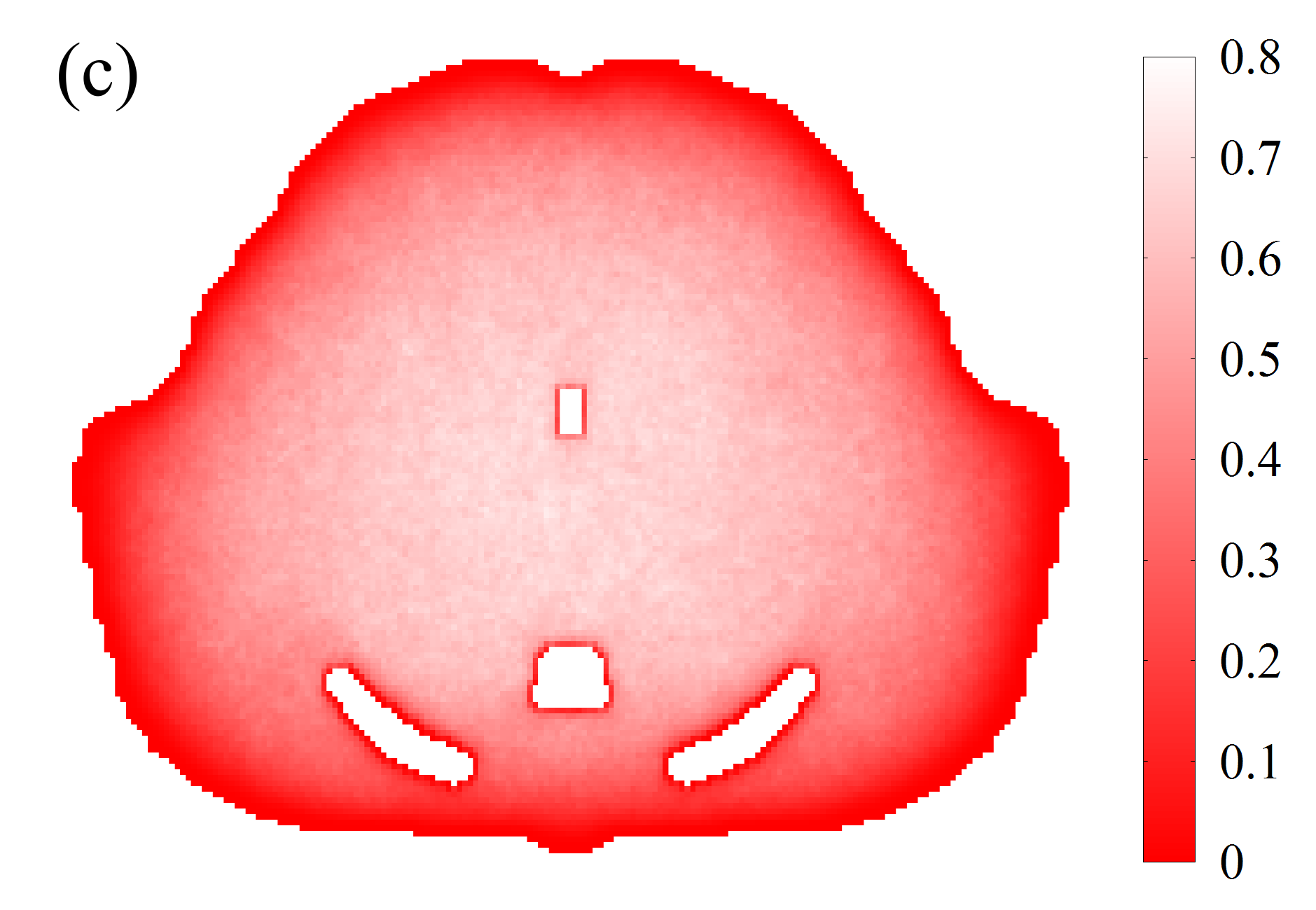}
\caption{Serotonergic fibers in a cross-section through the mouse midbrain.
(a) Fiber densities visualized with an anti-serotonin-transporter antibody. Higher densities are darker in the image; individual fibers are not visible at this magnification. Aq, cerebral aqueduct; cp, cerebral peduncle; IPN, interpeduncular nucleus; MGN, medial geniculate nucleus; ml, medial lemniscus; SC, superior colliculus; SN, substantia nigra; vtgx, ventral tegmental decussation. Scale bar = 1 mm.
(b) Single fiber modeled as superdiffusive FBM trajectory of $2^{17}$ steps with $\alpha=1.6$.
(c) Heat map of the simulated fiber density determined from 192 fibers of $2^{23}$ steps each,
plotted as ``optical'' density $\exp(-\beta P)$ with $\beta=38\,000$.
(The value of the attenuation parameter $\beta$ was chosen such that the mean pixel
value in the simulated section approximately matched the mean pixel value of the actual section in (a).) }
\label{fig:brain}
\end{figure}

Within this model, each individual serotonergic fiber is represented as the path of
a discrete FBM with a step size related to the thickness of the fibers (which determines how fast
the fibers can bend). A comparison of FBM sample paths with actual fiber trajectories suggests
that appropriate values of the anomalous diffusion exponent lie in the superdiffusive range.
Figure \ref{fig:brain} presents an example of a computer simulation of this model applied to
a section of a mouse brain.
The figure clearly shows that the simulations reproduce the increased fiber densities observed at the boundaries
of the real brain section as well as in the concave parts of its geometry.

A systematic study of this model and a detailed comparison with the fiber densities in real mouse brains
was carried out in Ref.\ \cite{JanusonisDeteringMetzlerVojta20}.
The agreement of the simulated  densities and the densities determined from the mouse brain sections
was found to be quite remarkable, especially in view of how little “neurobiological input” the model requires.
Moreover, the study demonstrated that “soft” fiber-wall interactions can be particularly appropriate for modeling
the behavior of serotonergic fibers in brain tissue.

\section{Conclusions}
\label{sec:conclusions}

In summary, we have performed large-scale computer simulations of FBM in one,
two, and three dimensions in the presence of reflecting boundaries that
confine the motion to finite regions in space. In all studied geometries,
we have found that the stationary probability density deviates strongly from
the flat distribution observed for normal Brownian motion. Specifically,
we have found particles to accumulate close to the reflecting walls for
superdiffusive FBM whereas they are depleted near the walls for subdiffusive
FBM.

This phenomenon is easy to understand qualitatively. If the correlations are
persistent (superdiffusive FBM), particles will attempt to continue in the
same direction upon reaching the wall and thus get trapped for a long time
\footnote{The probability of finding long periods of motion in predominantly
one direction is discussed in Ref.\ \cite{IbrahimBarghathiVojta14}
for power-law correlated disorder.}, increasing the probability density
near the wall. If the correlations are antipersistent (subdiffusive FBM),
 particles will preferably move away from the wall right after hitting it,
reducing the probability density at the wall. (We emphasize that the noise
correlations extend beyond the reflection events as the noise is externally
given.)

We note that these accretion and depletion effects arise from the
nonequilibrium nature of FBM. The fractional Langevin equation, which contains
the same fractional noise as FBM but fulfills the fluctuation-dissipation
theorem \cite{Kubo66}, reaches a thermal equilibrium stationary state that is
governed by the Boltzmann distribution. On a finite interval with reflecting
walls, this leads to a flat probability density, as was recently confirmed by
large-scale simulations \cite{VojtaSkinnerMetzler19} \footnote{The fractional
Langevin equation does show accretion and depletion effects, albeit weaker
ones, in nonstationary situations \cite{VojtaSkinnerMetzler19}.}. The
key role played by the fluctuation-dissipation theorem becomes clear if one
considers a generalized Langevin equation with long-time correlated fractional
Gaussian noise but instantaneous damping. For this equation, which violates
the fluctuation-dissipation theorem, simulations \cite{VojtaSkinnerMetzler19}
 have shown that the stationary probability density on a finite interval
is not uniform but resembles the corresponding result for FBM.

Our simulations have demonstrated, that the stationary probability density of
FBM on a finite one-dimensional interval features a power law singularity,
$P(x) \sim |x-w|^\kappa$, close to a reflecting wall at position $w$.
The exponent $\kappa$ follows the conjecture $\kappa = 2/\alpha -2$
\cite{WadaVojta18} with high accuracy. In higher dimensions, the stationary
probability density close to a smooth boundary (be it straight or curved)
 features a power singularity governed by the same exponent, $\kappa =
2/\alpha -2$, as in one dimension. Close to sharp (concave) corners, the
singularities are enhanced. When approaching the corner of a rectangle (along
a straight line), the probability density features a power law with exponent
$4/\alpha - 4$, and for a general $d$-dimensional orthotope (hyperrectangle),
the corresponding exponent is expected to read $2d/\alpha-2d$.

We emphasize that all our results are robust against changes in how the
reflecting walls are defined and implemented. In Sec.\ \ref{subsec:wall_test},
we have systematically compared simulations with four different types of
reflecting boundary conditions. These simulations demonstrate that details of
the wall implementation influence the probability density only in a narrow
``wall region'' whose size is determined by the step size $\sigma$ and becomes
unimportant in the continuum limit $L \gg \sigma$.  For soft walls, the size
 of the wall region is governed by the decay length of the wall force.

The nonuniform and singular probability density of reflected FBM can have
important consequences for applications. One such application is the modeling
of serotonergic fibers in the brain, as was discussed in
Sec.\ \ref{sec:fibers}. Here, the accretion and depletion of particles
close to reflecting walls and in concave parts of the geometry is crucial
for correctly describing variations of the experimentally observed fiber
densities in various brain regions. Note that active growth of these fibers
in the brain is clearly not an equilibrium process, supporting the use of FBM
rather than a fractional Langevin equation.

Recently, the logistic equation with temporal disorder, which describes the
evolution of a biological population density $\rho$ under environmental
fluctuation, was mapped onto FBM with a reflecting wall at the
origin \cite{VojtaHoyos15,WadaSmallVojta18}. This mapping relates the
density of individuals and the position of the walker through $\rho =
e^{-x}$. Consequently, the power-law singularity in the probability density
of FBM is intimately tied to the critical behavior of the nonequililibrium
phase transition between extinction and survival of the population and the
dependence of its universality class on the correlations in the environmental
fluctuations \cite{WadaSmallVojta18}.

In many realistic systems, the power-law correlations are regularized beyond
some time or length scale.  To account for such regularization effects on
the properties of confined FBM, one can employ tempered fractional Gaussian
noise \cite{MolinaGarciaetal18}.

Finally, we emphasize that the combination of  geometric confinement and
 long-time correlations provides a general route to a singular probability
density. We therefore expect analogous results for many long-range correlated
stochastic processes in nontrivial geometries.

\acknowledgments

This work was supported in part by a Cottrell SEED award from Research Corporation
and by the National Science Foundation under Grants No. DMR-1828489 and No.
OAC-1919789 (T.V.).
S.J. acknowledges support by the National Science Foundation
under Grant Nos. 1822517 and 1921515, by the National Institute of Mental Health (Grant No. MH117488), and
by the California NanoSystems Institute (Challenge-Program Development grant).
R.M. acknowledges support from the German Research Foundation (DFG, Grant No. ME/1535/7-1)
and from the Foundation for Polish Science (Fundacja na rzecz Nauki Polskiej,
FNP) within an Alexander von Humboldt Polish Honorary Research Scholarship.

\bibliographystyle{apsrev4-1}
\bibliography{../00Bibtex/rareregions}
\end{document}